\documentclass[aps,prb,preprint,groupedaddress,amsmath,amssymb]{revtex4}

\usepackage{graphicx}
\usepackage{dcolumn}
\usepackage{bm}
\usepackage{hyperref}

\begin{document}

\title{Magnetic order in van der Waals antiferromagnet CrPS$_4$: anisotropic $H - T$ phase diagrams and effects of pressure}

\author{ Sergey L. Bud'ko$^{1,2}$, Elena Gati$^{1,2}$, Tyler J. Slade$^{1,2}$,
and  Paul C. Canfield$^{1,2}$}

\affiliation{$^{1}$Ames Laboratory, US DOE, Iowa State University, Ames, Iowa 50011, USA}
\affiliation{$^{2}$Department of Physics and Astronomy, Iowa State University, Ames, Iowa 50011, USA}

\date{\today}

\begin{abstract}

Single crystalline samples of the van der Waals antiferromagnet CrPS$_4$ were studied by measurements of specific heat and comprehensive anisotropic temperature- and magnetic field- dependent magnetization. In addition, measurements of the heat capacity and magnetization were performed under pressures of up to $\sim 21$~kbar and $\sim 14$~kbar respectively. At ambient pressure, two magnetic transitions are observed,  second order from a paramagnetic to an antiferromagnetic state at $T_N \sim 37$ K, and a first order spin reorientation transition at $T^* \sim 34$ K.  Anisotropic  $H - T$ phase diagrams were constructed using the $M(T,H)$ data. As pressure is increased, $T_N$ is weakly suppressed with $dT_N/dP \approx - 0.1$~K/kbar. $T^*$, on the other hand, is suppressed quite rapidly, with  $dT^*/dP \approx - 2$~K/kbar, extrapolating to a possible quantum phase transition at $P_c \sim 15$~ kbar.

\end{abstract}



\maketitle

\section{Introduction}

Two-dimensional (2D) van der Waals (vdW) materials,  in which weak interactions between the 2D layers allow the crystals to be readily cleaved or exfoliated, have attracted significant attention due to their novel properties often associated with reduced dimensionality.\cite{bhi15a,mcg20a} These materials  open a wide range of possibilities for fundamental research and applications.  They offer new means to study 2D magnetism, where spin fluctuations are expected to be strongly enhanced. \cite{sam17a}  Furthermore, novel exotic quantum phases,  including the quantum Hall effect, quantum spin Hall effect, and quantum spin liquid, are expected to be observed in these materials and related heterostructures. \cite{par16a,bur18a} Two-dimensional vdW materials also allow for control and manipulation of magnetic states through coupling to external perturbations such as strain, magnetic and electric fields, and moir\'e patterns. \cite{bur18a}  For applications, 2D magnetism, combined with semiconductivity, would offer the path to spintronic devices. \cite{and06a,bal18a} 

So far, most studies on vdW magnetic materials are focused on ferromagnets (e.g., Cr$_2$Ge$_2$Te$_6$, CrI$_3$ etc. \cite{hua17a,gon17a}), with less attention paid to 2D materials with antiferromagnetic (AFM) order. The semiconducting chromium thiophosphate, CrPS$_4$, belongs to this less explored class of 2D vdW antiferromagnets. It was first reported more than four decades ago \cite{tof77a,die77a} as crystallizing in a layered,  monoclinic structure (space group {\it C2/m} \cite{tof77a} or {\it C2} \cite{die77a}) that has a clear vdW gap between the CrPS$_4$ layers. These layers lie within the $ab$ basal plane with the $c$-axis making angle $\beta \approx 91.9^{\circ}$, close to 90$^{\circ}$, with respect to the $a$-axis. Early anisotropic magnetic susceptibility measurements \cite{lou78a} indicated antiferromagnetic ordering at $T_N \approx 36$ K with the moments confined to the basal, $ab$ plane.

Interest in CrPS$_4$ was recently renewed, \cite{pei16a} and magnetization measurements confirmed the earlier results. \cite{lou78a} Furthermore, a metamagnetic transition was observed for $H \| c$, and a schematic low field ($H \leq 10$~kOe) $H - T$ phase diagram for this field direction was constructed. Based just on magnetization measurements, it was suggested that below the $T_N$,  in  zero magnetic field, CrPS$_4$ is a C - type AFM that transforms into a G - type AFM as a result of the metamagnetic transition upon application of the magnetic field in the $c$ - direction. The electrical resistivity was reported to have semiconducting behavior with an energy gap of $E_a = 0.166$ eV, \cite{pei16a} whereas the optical measurements reported the gap values of $\sim 1.3-1.4$ eV \cite{lee17a,sus20a} or $\sim 2.4$ eV. \cite{lou78a}  Heat capacity measurements confirmed $T_N \approx 36$ K with the evaluated magnetic entropy being consistent with $S = 3/2$ state of the Cr$^{3+}$ ion. 

Using a combination of bulk magnetization, torque measurements \cite{pen20a,din20a} and powder neutron diffraction on a pressed pellet of powders (microflakes) in zero and at several applied magnetic fields, \cite{pen20a} an extended $H - T$ phase diagram for $H \| c$ was reported. The low temperature AFM state was identified as A - type formed by ferromagnetic Cr layers (with the magnetic moments slightly $\approx 9.5^\circ$ tilted from the $c$ - axis toward the $a$ - axis) that are ordered antiferromagnetically along the $c$ - axis. As the result of the metamagnetic transition (for $H \| c$), the order was reported to change to a canted AFM with the magnetic moments aligned close to the $b$ - direction. These magnetic structures are at odds with those proposed in Ref. \onlinecite{pei16a} based on magnetization data.

Comprehensive elastic and inelastic neutron diffraction measurements on single crystals and powder of CrPS$_4$ were performed in Ref. \onlinecite{cal20a}.  In this publication, the $H = 0$, low temperature magnetic structure was described as ferromagnetic sheets of Cr moments aligned primarily along the $c$ - axis with a small component along the $a$ - direction. These sheets are ordered antiferromagnetically along the $c$ - direction. As a result of the spin - flop transition for the  magnetic field applied in the $c$ - direction, the moments were reported to rotate and align along the $b$ - direction, followed by subsequent gradual rotation towards $c$,  the applied field direction, on further field increase. Overall interpretation of the neutron diffraction data is similar to the results of Ref. \onlinecite{pen20a}.  Inelastic neutron scattering results were modeled with four  (three in-plane and one out-of-plane) exchange interactions ranging from $\sim -3$ meV to $\sim 0.2$ meV.\\

Whereas the nature of the low temperature magnetic structures is now fairly well understood, the behavior of the magnetic susceptibility in the 30 - 50 K region is rather complex with an abnormal minimum at $\sim 35$ K for $H \perp c$. \cite{pei16a,pen20a,cal20a} The interpretation of this feature is not established and was suggested to be induced by in-plane short-range ferromagnetic correlations competing with out-of-plane AFM ordering \cite{pen20a} or to be a signature of a subtle spin reorientation transition.\cite{cal20a}. One of the goals of this work is to study of the magnetic ordering in CrPS$_4$ using heat capacity as a complementary technique, as well as detailed magnetization measurements and analysis to construct anisotropic $H - T$ phase diagrams over the whole $T \leq T_N$ range.\\

Another goal of this work is to study effect of pressure on magnetism in CrPS$_4$.  Understanding the effects of stress on magnetism is of particular importance for materials with potential applications in monolayer form or in heterostructures.\cite{lee17a,lee18a,kim19a,bud20a} Indeed, first principle calculations \cite{zhu16a,joe17a} suggested the magnetic order changes to ferromagnetic in monolayer and stressed monolayer samples of CrPS$_4$. The properties of bulk materials studied under hydrostatic pressure often serve as an experimental benchmark to validate predictions of band structure calculations, sometimes yielding a deeper insight into material properties. \cite{gat19a} A study of the effect of hydrostatic pressure on magnetism in CrPS$_4$ is a second goal of this work that is achieved using two thermodynamic probes, magnetization and ac calorimetry, up to $\sim 14$ and $\sim 21$ kbar, respectively. The only high pressure study of CrPS$_4$\cite{sus20a} we are aware of presents an observation of direct to indirect band gap crossover and an insulator–metal transition under pressure using optical and electrical transport measurements and does not address any magnetic properties of this material.

\section{Experimental details}

CrPS$_4$ single crystals were grown by chemical vapor transport (CVT).\cite{pen20a} Chromium powder (Alfa Aesar, 99\%), red phosphorus pieces (Alfa Aesar, 99.999\%), and sulfur pieces (Alfa Aesar, 99.999\%) were weighed and mixed in a 1:1:4 molar ratio.  The elements were placed in an amorphous silica tube with elemental iodine to serve as a transport agent.  The tube ($\sim 18$ cm long, $\sim 1.5$ cm diameter) was flame sealed under vacuum and placed horizontally into a two-zone furnace.  Both zones of the furnace were simultaneously heated to 200$^{\circ}$C over 10 h, then to 450$^{\circ}$C over 24 h, and finally to 600$^{\circ}$C over an additional 24 h.  Prior to crystal growth, in order to minimize the number of nucleation sites on the growth side, the growth side was heated to 680$^{\circ}$C over 3 h and allowed to dwell for 24 h, while the temperature of source end, containing the elements, was held at 600$^{\circ}$C.  For the crystal growth, the source end was brought to 680$^{\circ}$C and the growth side of the ampoule was brought  to 600$^{\circ}$C.  The furnace was held under these conditions for 192 h after which the tube was slowly removed from the furnace to reveal dark, blade-like crystals of CrPS$_4$ of up to $10 \times 2 \times 0.5$~mm$^3$ size on the growth side and near the center of the tube.

To confirm the phase of the CVT grown crystals, samples were analyzed with powder X-ray diffraction.  Several crystals were hand ground to a powder and diffraction patterns were collected on a Rigaku Miniflex-II instrument operating with Cu–K$\alpha$ radiation ($\lambda = 1.5406$ \AA) at 30 kV and 15 mA. The pattern was fit using Rietveld refinement as implemented in GSAS-II software package \cite{tob13a} The diffraction pattern is consistent with the published crystal structure and lattice parameters (see the Appendix for the powder x-ray data as well as the details and results of the structure refinement). 

\begin{figure}
\begin{center}
\includegraphics[angle=0,width=140mm]{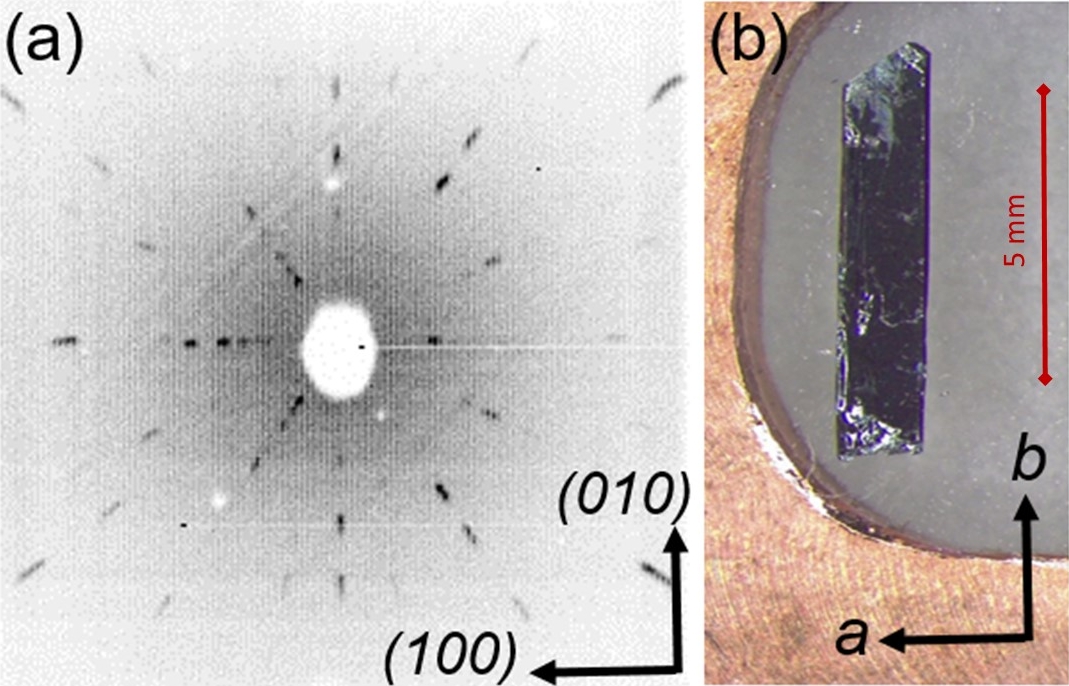}
\end{center}
\caption{(Color online) (a)  Laue pattern, with in-plane crystallographic directions marked, of the CrPS$_4$ crystal; (b) the crystal  mounted on the sample holder.} \label{F0}
\end{figure}

The back-scattering Laue imaging/pattern of the crystal [Fig. \ref{F0}(a)] was obtained with a Multiwire Laue Camera using x-ray from a Mo x-ray tube  ($\lambda = 0.7107$ \AA) operated at 8.5 V and 30 mA. The sample was mounted on a goniometer using double-sided Scotch tape [Fig. \ref{F0}(b)]. The measurement was conducted with 10 cm between the sample and detector and the beam perpendicular to the plate-like surface of the crystal. Based on Laue measurements [Fig. \ref{F0}(a)] for these crystals, the  $c^*$-axis is perpendicular to the plate-like surface, the $b$-axis is along the longer direction of the blade and the $a$-axis is in-plane, perpendicular to the longer direction of the blade. Since the angle $\beta$ was reported to be $91.88^{\circ}$, very close to 90$^{\circ}$, in  the rest of the text we will use the notation of the $c$-axis, or $c$-direction instead of $c^*$ for the direction perpendicular to the plate-like surface. Generally speaking, the accuracy of the orientation of the applied field with respect to a given crystallographic axis in our experiments is not better than $5^{\circ}$.

Ambient pressure magnetization measurements were performed as a function of temperature (1.8 - 300 K) and magnetic field (up to 70 kOe) using a vibrating sample magnetometer (VSM) option of the Quantum Design MPMS 3 Magnetic Property Measurement System. For $H \| a$ and $H \| b$ measurements, the sample was mounted on a fused silica sample holder, whereas a brass sample holder was used for  $H \| c$ measurements. The data were not corrected for the (very small) sample holders' background signal or contributions due to the sample's finite dimensions. \cite{qd14a} For the samples and holders used in this work the total effect of both contributions is, conservatively, less than 10\% and has no bearing on the position of the magnetic transitions.  Specific heat at ambient pressure was measured using a hybrid adiabatic relaxation technique of the heat capacity option in a Quantum Design Physical Property Measurement System (PPMS) instrument.

Magnetization measurements under pressure were performed using a commercial HMD piston-cylinder pressure cell \cite{hmd} with  Daphne oil 7373 (solidification pressure of $\sim 20$ kbar at 300 K \cite{tor15a}) as a pressure medium. Three separate pressure runs up to $\sim 14$ kbar were performed. In all runs, the sample was oriented with the applied field ($H = 1$ kOe) approximately along the $b$-axis. The superconducting transition of  elemental Pb was used to determine pressure. \cite{eil81a}

Both at ambient pressure and under pressure the longitudinal component of magnetization, $M^{\|H}$, was measured.

Specific heat measurements under pressure up to $\sim 21$ kbar were performed using an ac calorimetry technique. A Be-Cu/Ni-Cr-Al hybrid piston-cylinder cell, similar to the one described in Ref. \onlinecite{bud86a}, was used. A 40:60 mixture of light mineral oil:n-pentane, which solidifies at room temperature at $\sim 35$ kbar, \cite{tor15a}  was used as a pressure medium. Elemental Pb was used as a low temperature pressure gauge. \cite{eil81a}  Quantum Design PPMS was used to provide the temperature environment. Details of the setup used and the measurements protocol are described in Ref. \onlinecite{gat19b}.

\section{Results and Discussion}

\subsection{Magnetic order and anisotropic $H - T$ phase diagrams}
\begin{figure}
\begin{center}
\includegraphics[angle=0,width=140mm]{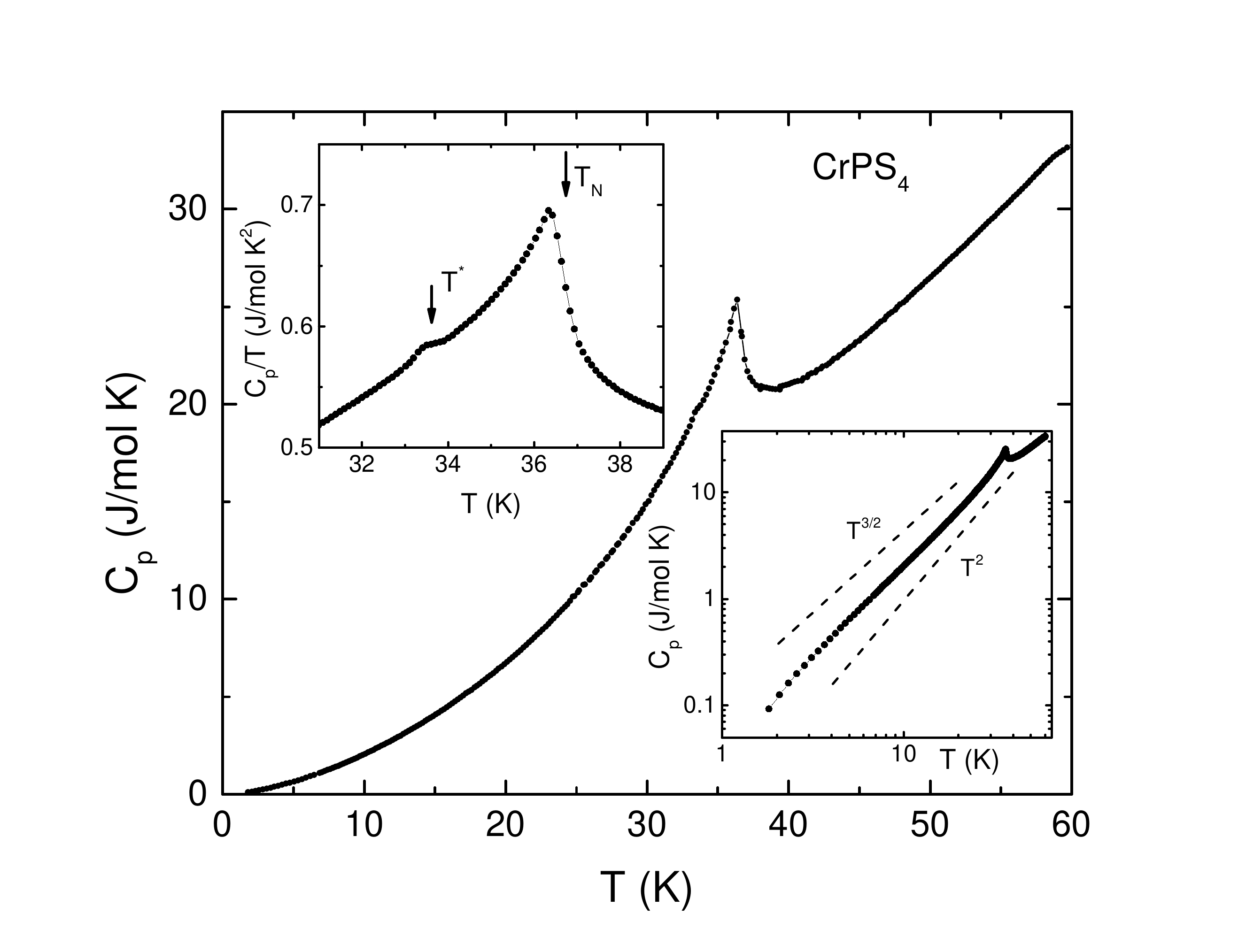}
\end{center}
\caption{Low temperature, ambient pressure, zero magnetic field, specific heat of CrPS$_4$. Upper inset: enlarged view of specific heat data set taken in the vicinity of the transitions and plotted as $C_p/T$ vs. $T$. Arrows mark the positions of the transitions. The transition temperatures were determined using the position of the minima in $d(C_p/T)/dT$ as a criterion. This criterion is close to the standard criterion that is using iso-entropic construction. Lower inset: $C_p(T)$ data of the main panel plotted on a log-log scale. Dashed lines show $T^{3/2}$ and  $T^2$ functional behavior.} \label{F1}
\end{figure}

The low temperature specific heat of CrPS$_4$ at ambient pressure is shown in Fig. \ref{F1}.  In addition to the dominant feature at $T_N = 36.7$ K, another smaller feature at $T^* = 33.7$ K is clearly seen. This suggests that the lower temperature transition is bulk but with the associated magnetic entropy is estimated to be just few percent of the magnetic entropy of the dominant transition. This is consistent with $T^*$ being a spin reorientation transition. The evidence for the $T^*$ transition in magnetization data will be discussed below.

We would like to mention that at low temperatures, in the magnetically ordered state, the specific heat of CrPS$_4$ approximately follows $T^{\alpha}$ behavior where the $\alpha$ value is in the range between 1.5 and 2  (see Fig. \ref{F1}, lower inset). This is different from the $T^3$ behavior expected both for the  phonon contribution in three-dimensional materials or the power law for the magnon contribution in simple antiferromagnets. \cite{gop66a} A value of $\alpha = 2$ in some range of low temperatures can be obtained in simple models of the phonon contribution in 2D solids \cite{tar45a,gur52a} and is observed  in graphite and other 2D materials. \cite{des53a,its57a} However, the magnon contribution to the specific heat  in real, complex materials is defined by the magnon spectrum and can deviate from what is expected from simple models. Attempts at modeling the magnon contribution to the specific heat based on inelastic neutron scattering data \cite{cal20a} as well as realistic theoretical evaluation of the phonon contribution  are complex and fall beyond the scope of this work. All in all, our data appear to be consistent with the 2D character of CrPS$_4$.

\begin{figure}
\begin{center}
\includegraphics[angle=0,width=115mm]{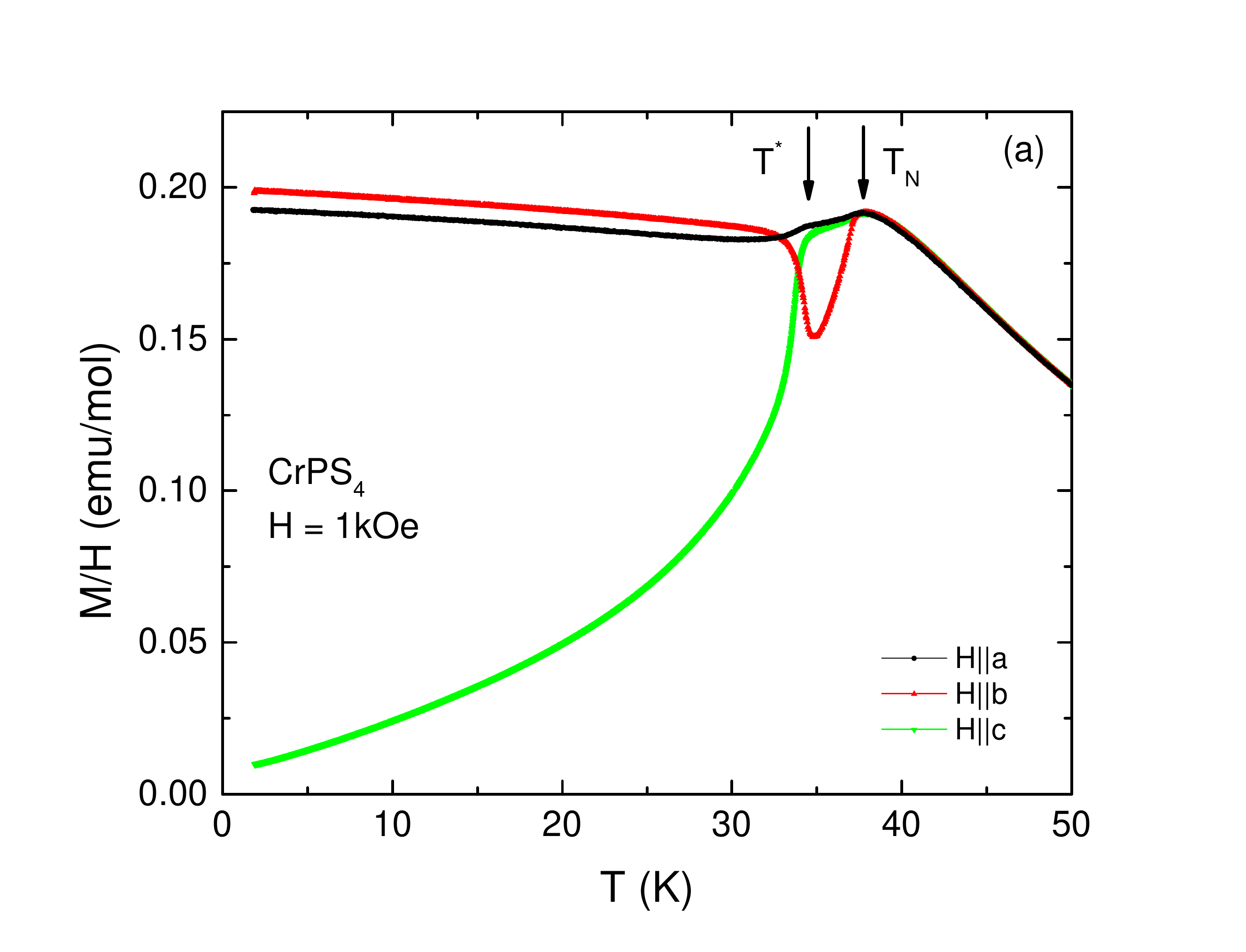}
\includegraphics[angle=0,width=115mm]{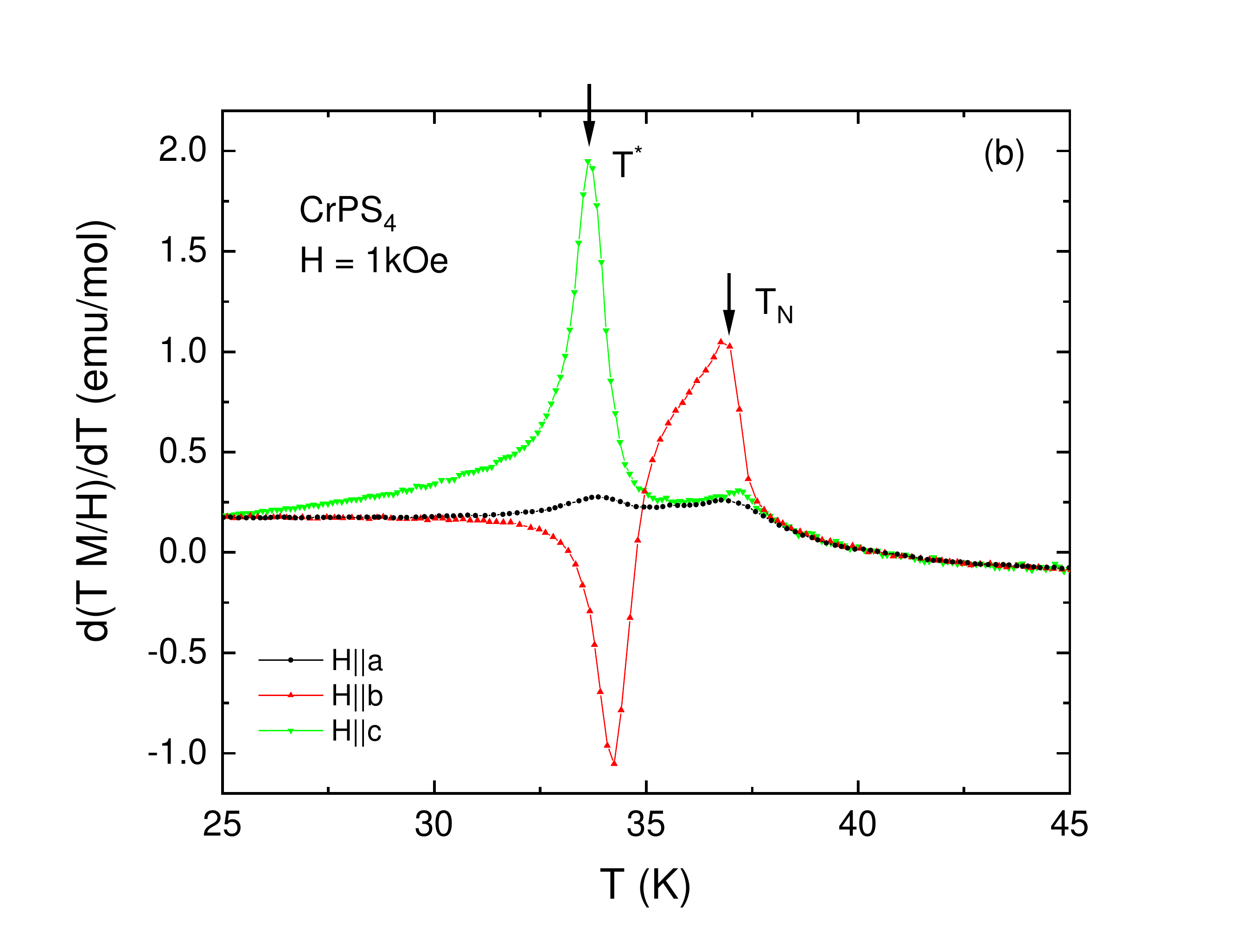}
\end{center}
\caption{(Color online) (a) Anisotropic low temperature magnetic susceptibility, $M/H$ vs $T$, of CrPS$_4$. Note, the data for $H \| a$ and $H \| c$ were normalized to match the $H \| b$ data at 50 K to account for small sample shape and mounting effects in VSM measurements using MPMS 3. (b) Temperature derivatives $d(\chi T)/dT$ for three directions of the magnetic field shown in the vicinity of the transitions. Arrows mark the positions of the transitions. Reduction by group data analysis feature, causing reduction in data density, was used before calculating the derivatives.} \label{F2}
\end{figure}

The anisotropic, low temperature magnetic susceptibility of CrPS$_4$ is shown in Fig. \ref{F2}(a). In the picture of a simple, easy-axis antiferromagnet, \cite{blu01a} the data below $T^*$ are consistent with the moments aligned close to the $c$ - axis, whereas the data between $T_N$ and $T^*$ are consistent with the moments along $b$. This change from moments $\sim$ along the $c$-axis to moments $\sim$ along the $b$-axis on warming implies that close to $T^*$ these two moment directions (and structures) are very similar in energy. The transition temperatures were determined from the extrema of the derivatives $d(\chi T)/dT, \chi = M/H$ \cite{fis62a}. Indeed, there are two clear features for each of the three orientations [see Fig. \ref{F2}(b)], yielding the transition temperatures  $T_N = 36.9$ K, $T^* = 34.0$ K for $H \| a$, 36.8 K and 34.2 K for $H \| b$, 37.2 K, 33.6 K for $H \| c$,  resulting in average values of  $T_N = 37.0$ K, $T^* = 33.9$ K, consistent with those determined from the specific heat.

\begin{figure}
\begin{center}
\includegraphics[angle=0,width=140mm]{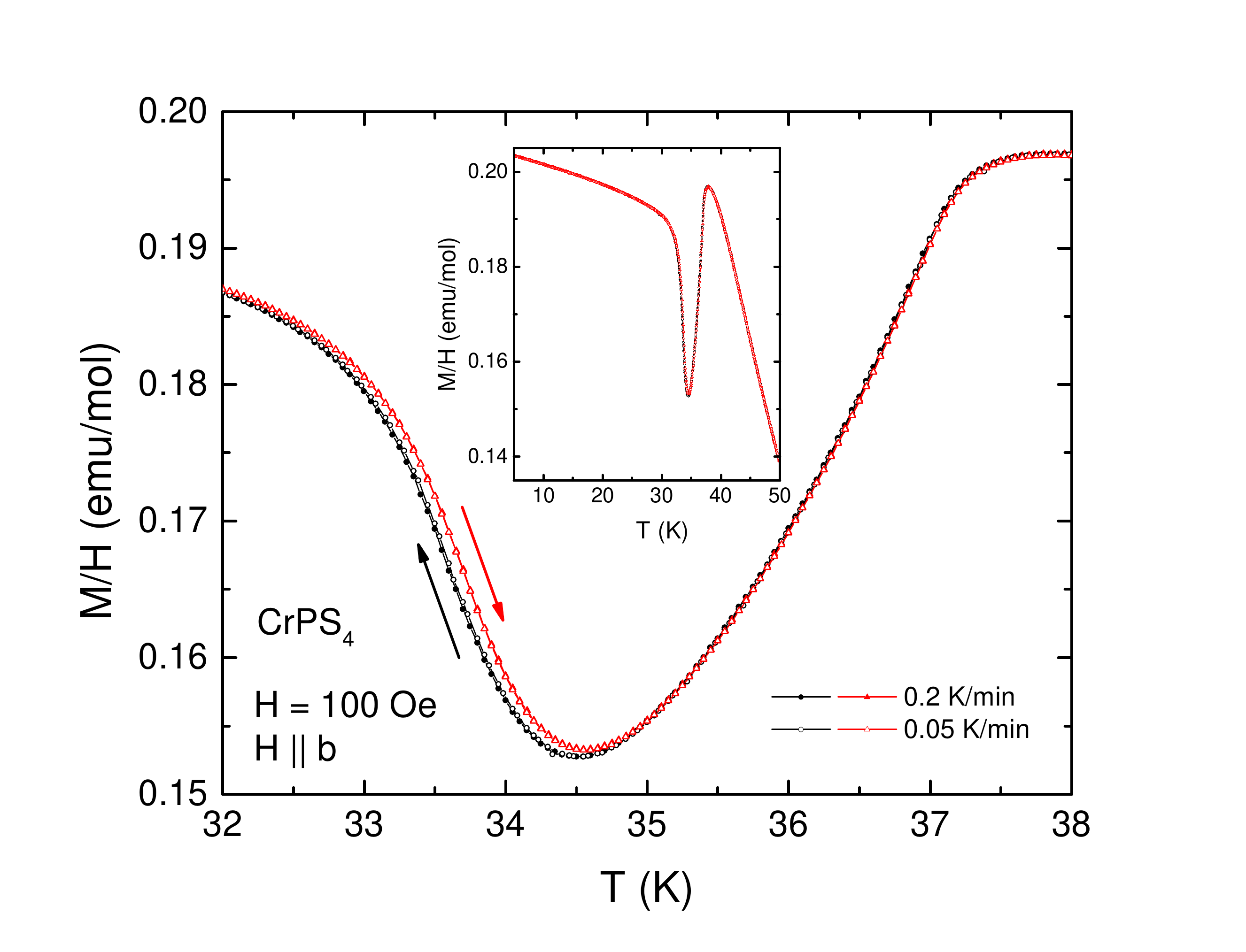}
\end{center}
\caption{(Color online) Temperature dependent suseptibility, $M/H$, measured at $H = 100$~Oe with $H \| b$ on cooling and warming in the vicinity of the magnetic transitions. Inset - the same data on a larger temperature scale. Measurements were performed with two different temperature sweep rates. } \label{FA2}
\end{figure}

To explore the order of the magnetic transitions, we measured low field ($H = 100$~Oe) temperature dependent magnetization for $H \| b$ on heating and cooling, see Fig. \ref{FA2}. The lower transition, $T^*$, shows clear, even though small, $\Delta T \approx 0.1$~K, hysteresis that does not change even with a factor of four change in the temperature sweep range ($dT/dt$). These data suggest that the $T^*$ transition is of the first order. There was no hysteresis observed for $T_N$, suggesting that this transition is likely second order. It should be noted that given that the $T^*$ transition is the first order, the use of the criterion based on  $d(\chi T)/dT$ may not formally be correct \cite{fis62a}, but for consistency it is used for all $M(T)$ data.
\\

\begin{figure}
\begin{center}
\includegraphics[angle=0,width=120mm]{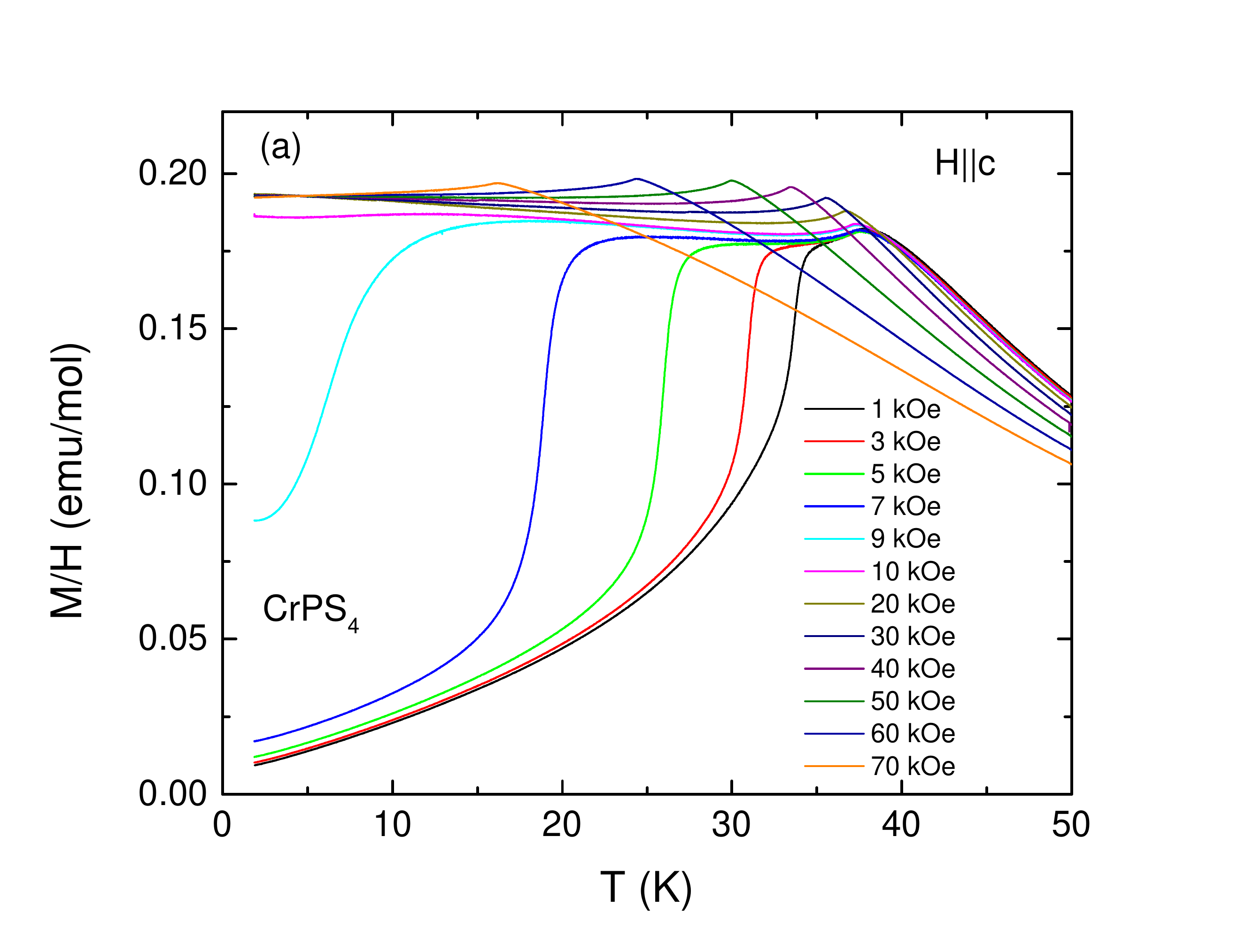}
\includegraphics[angle=0,width=120mm]{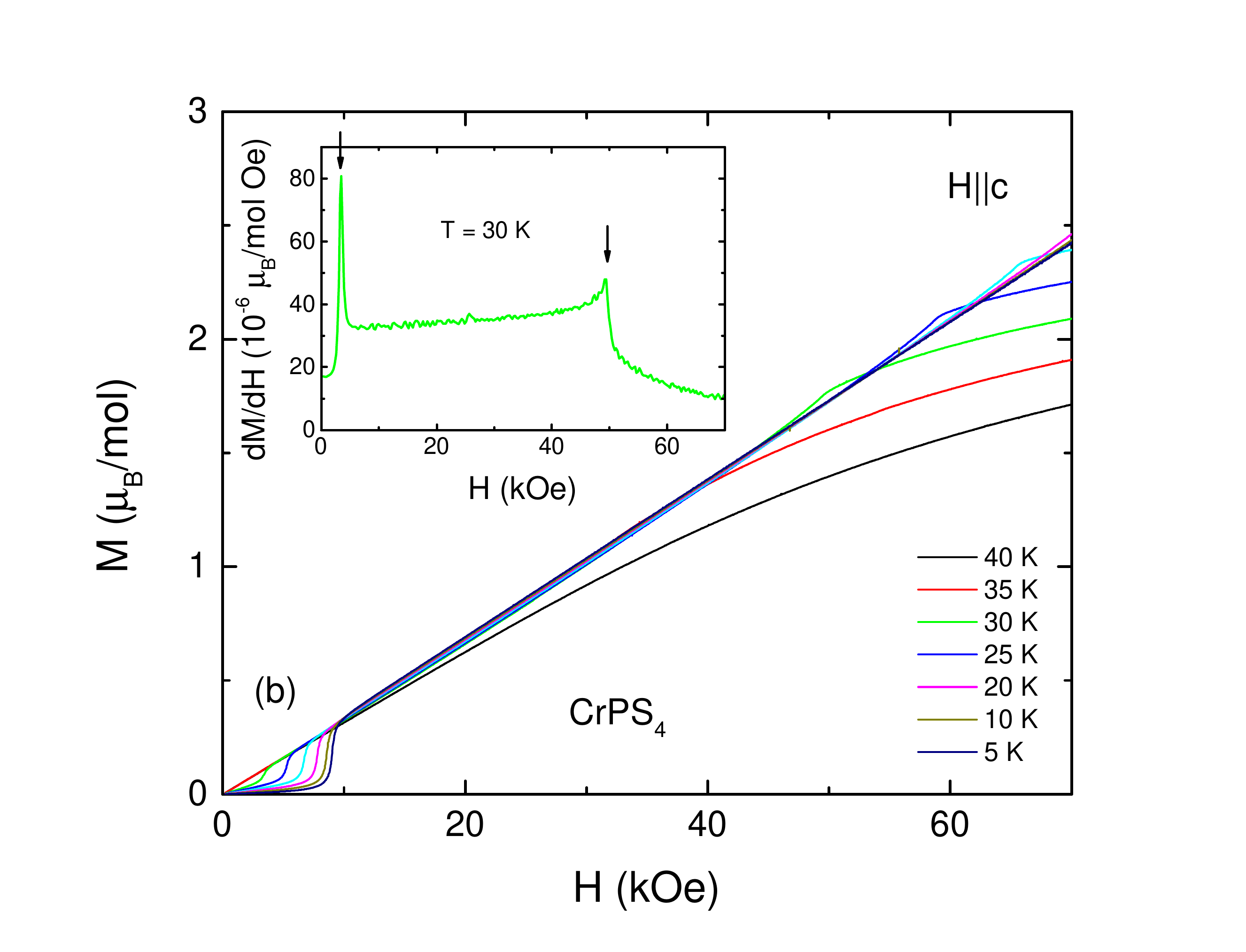}
\end{center}
\caption{(Color online) Selected examples of (a) $M(T)$ and (b) $M(H)$ measurements for $H \| c$. Inset in the panel (b) shows the derivative $dM/dH$ vs $H$ for 30 K $M(H)$ data. Arrows in the inset mark the positions of the transitions.} \label{F3}
\end{figure}

\begin{figure}
\begin{center}
\includegraphics[angle=0,width=120mm]{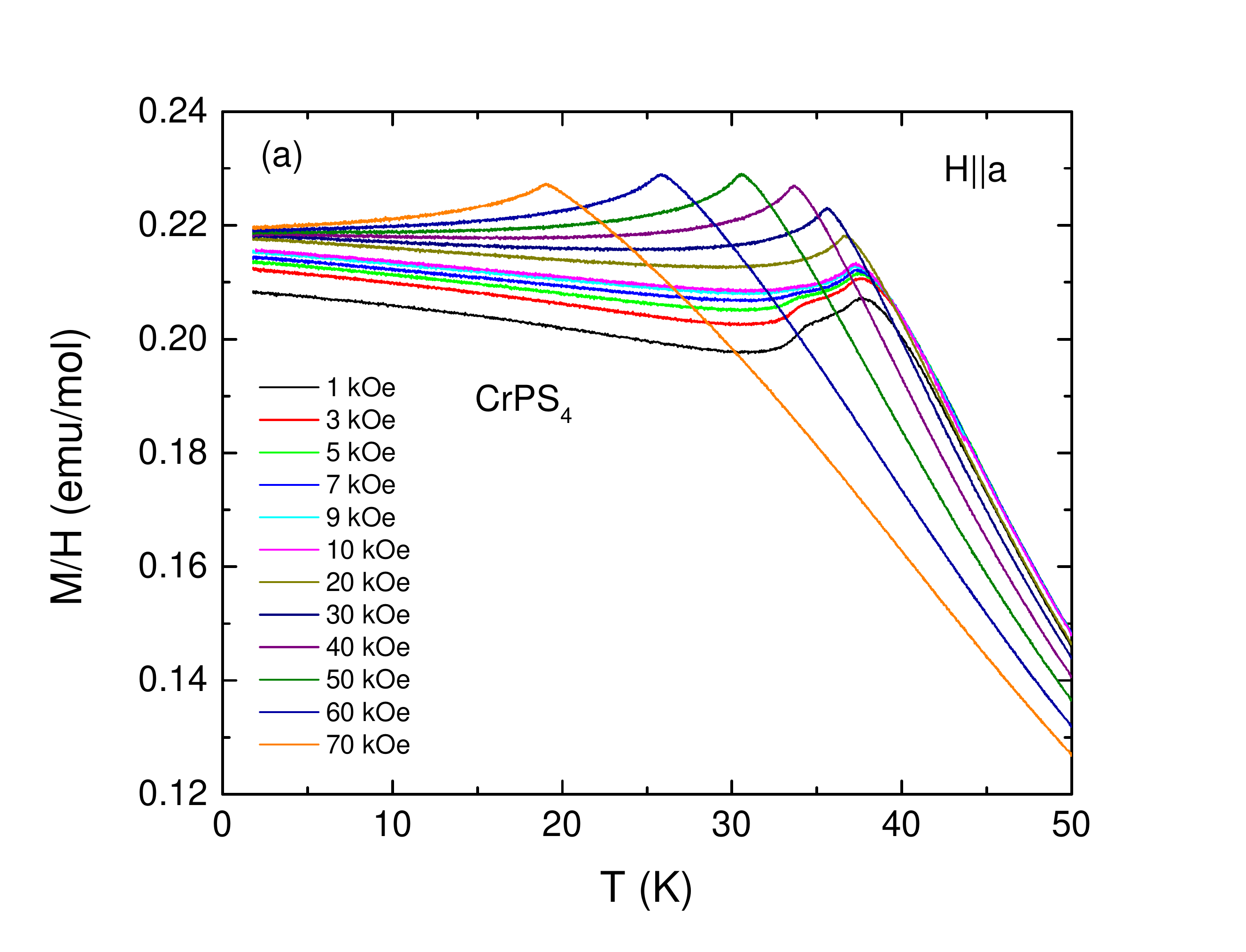}
\includegraphics[angle=0,width=120mm]{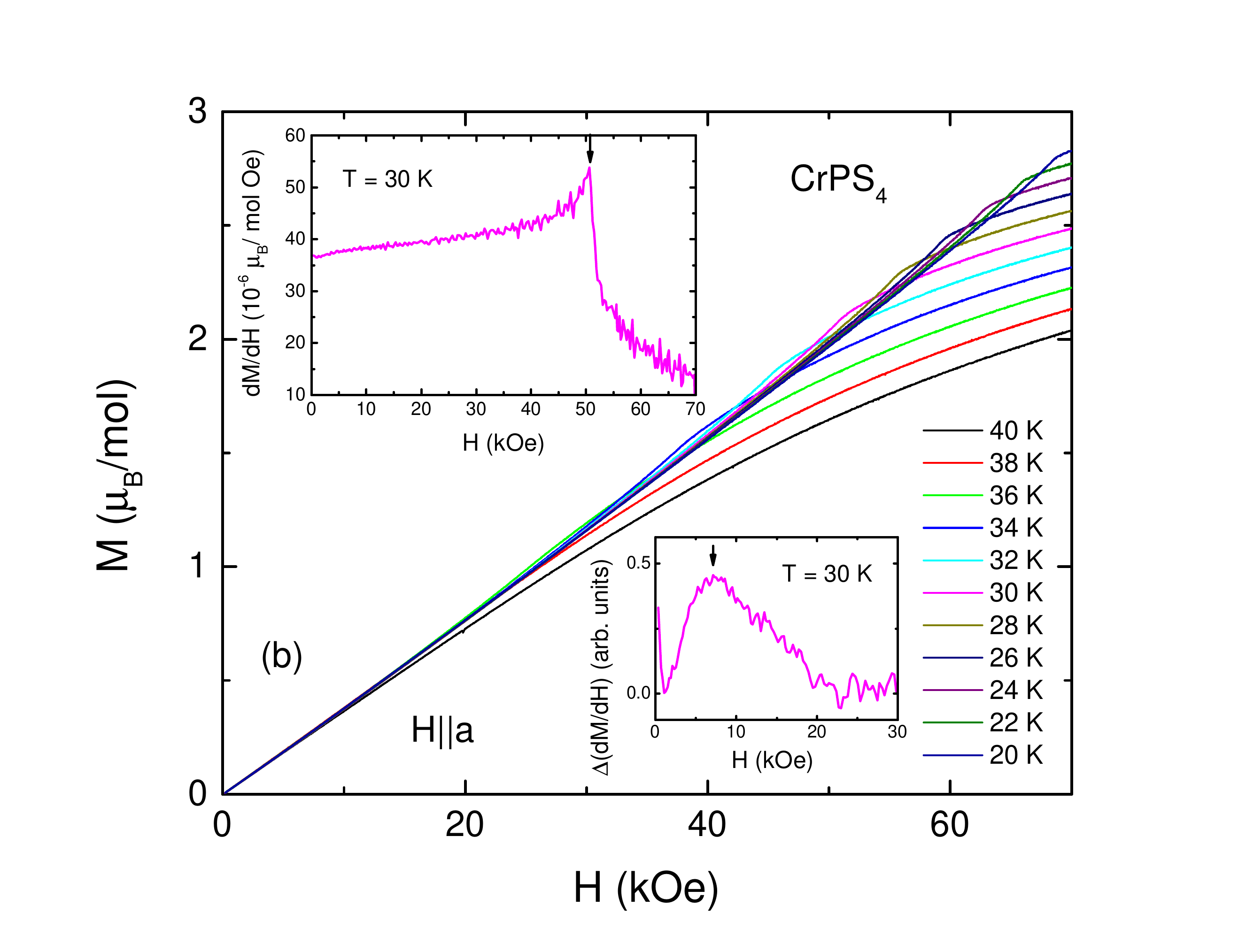}
\end{center}
\caption{(Color online) Selected examples of (a) $M(T)$ and (b) $M(H)$ measurements for $H \| a$. Insets in the panel (b) show the derivatives $dM/dH$ vs $H$ for 30 K $M(H)$ data. For the  lower right inset (low field part of the data), the linear background from the derivative was subtracted.   Arrows in the insets mark the positions of the transitions.} \label{F4}
\end{figure}

\begin{figure}
\begin{center}
\includegraphics[angle=0,width=120mm]{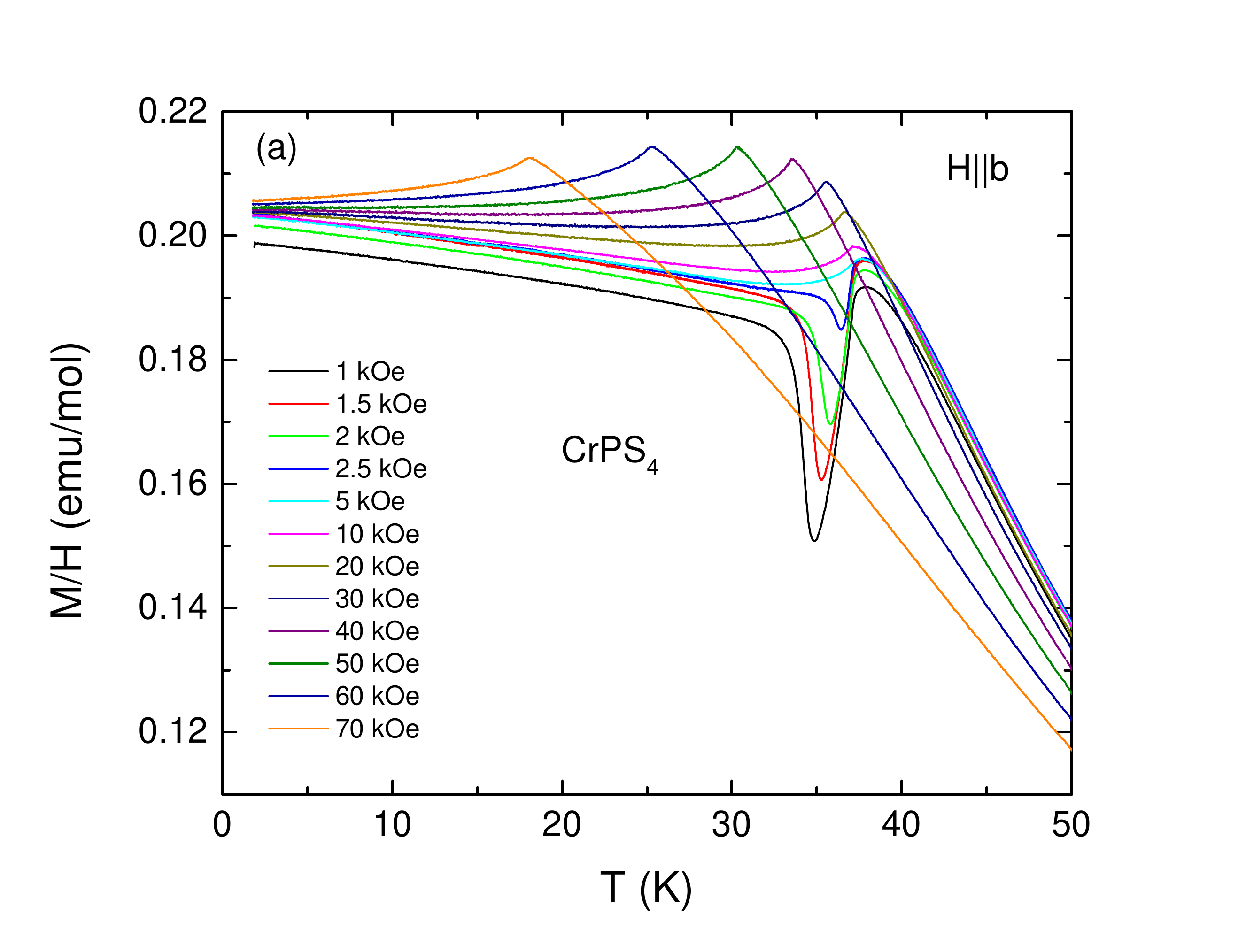}
\includegraphics[angle=0,width=120mm]{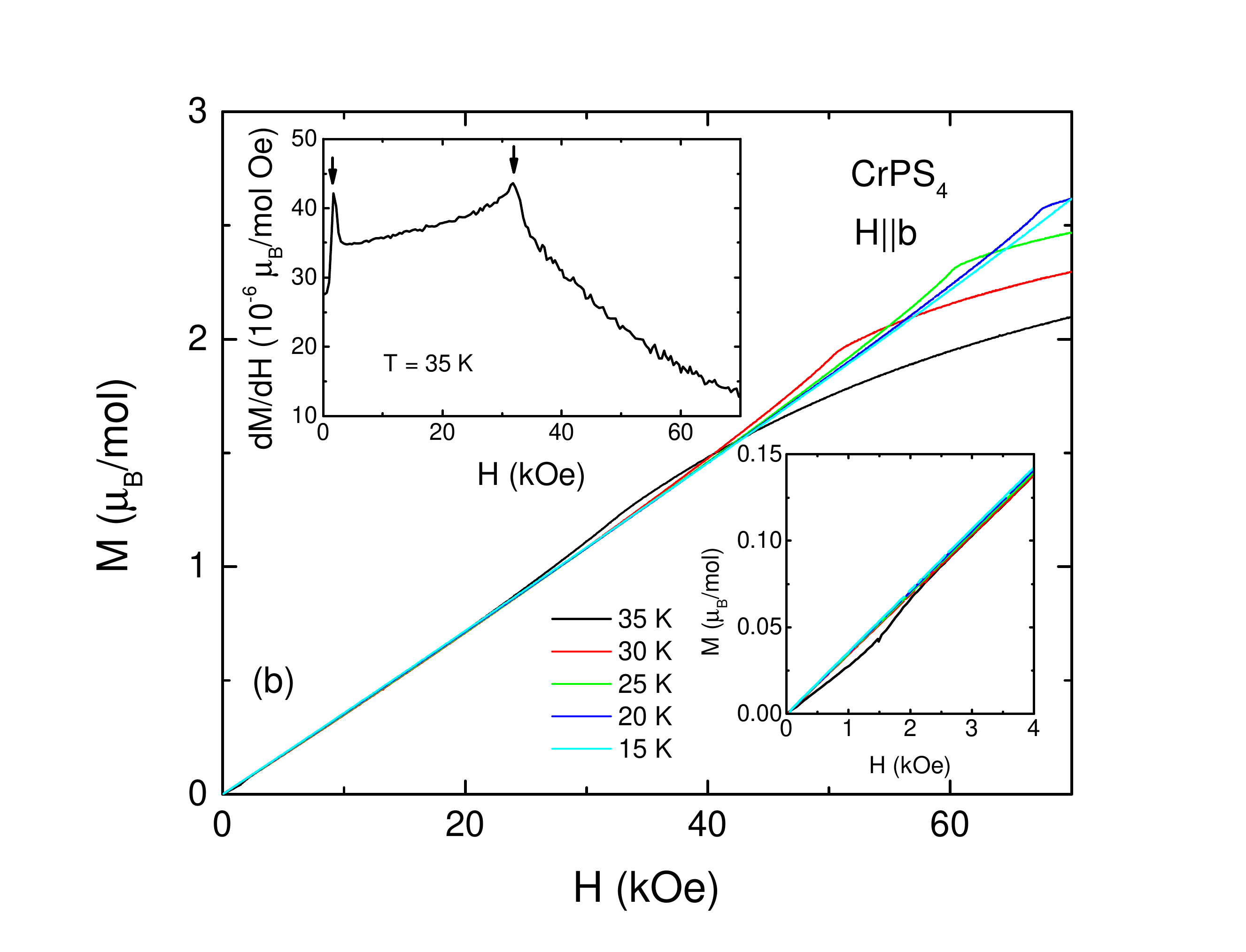}
\end{center}
\caption{(Color online) Selected examples of (a) $M(T)$ and (b) $M(H)$ measurements for $H \| b$. Lower inset in the panel (b) shows low field part of the data, upper inset shows the derivative $dM/dH$ vs $H$ for 35 K $M(H)$ data. Arrows in the inset mark the positions of the transitions.} \label{F5}
\end{figure}

\begin{figure}
\begin{center}
\includegraphics[angle=0,width=80mm]{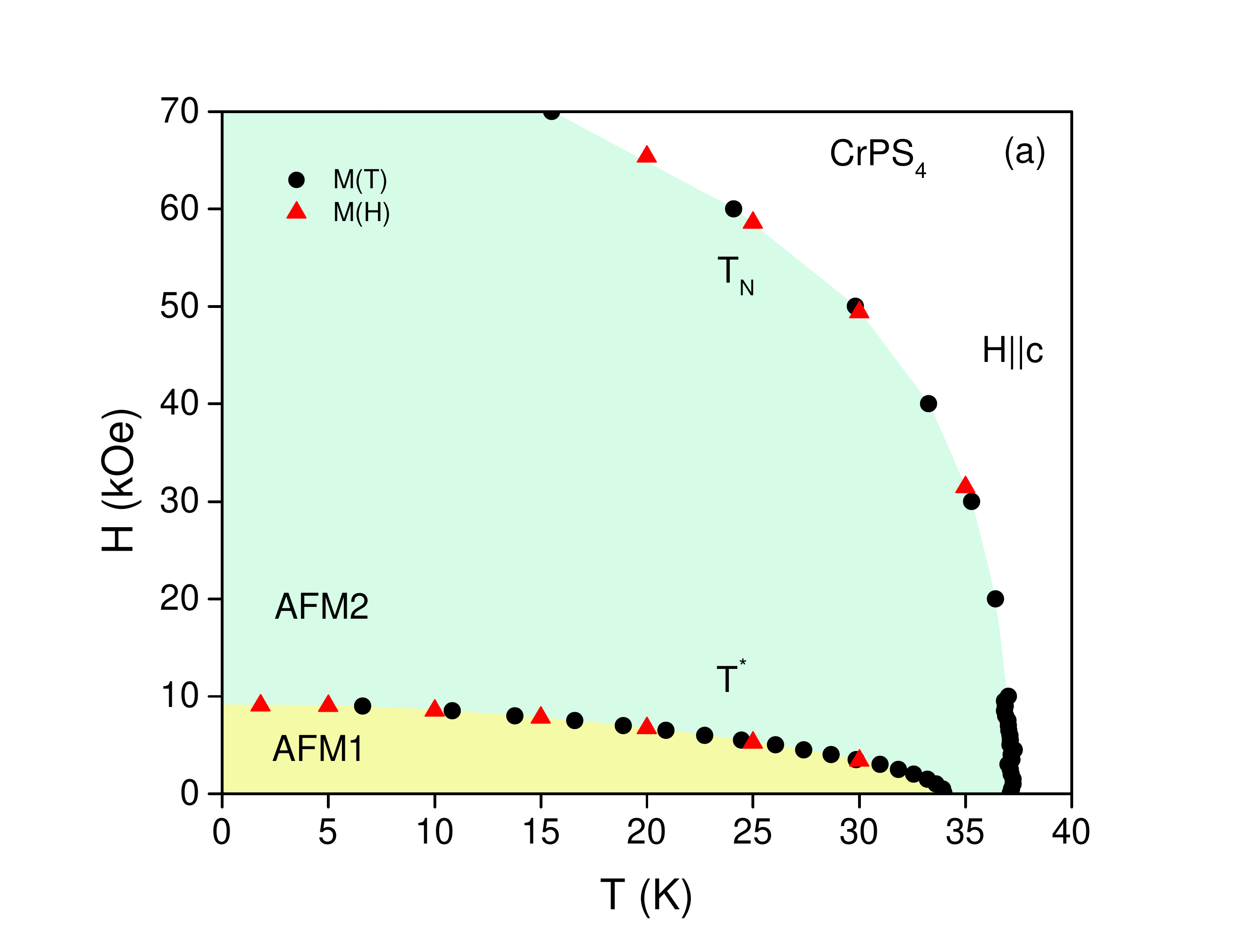}
\includegraphics[angle=0,width=80mm]{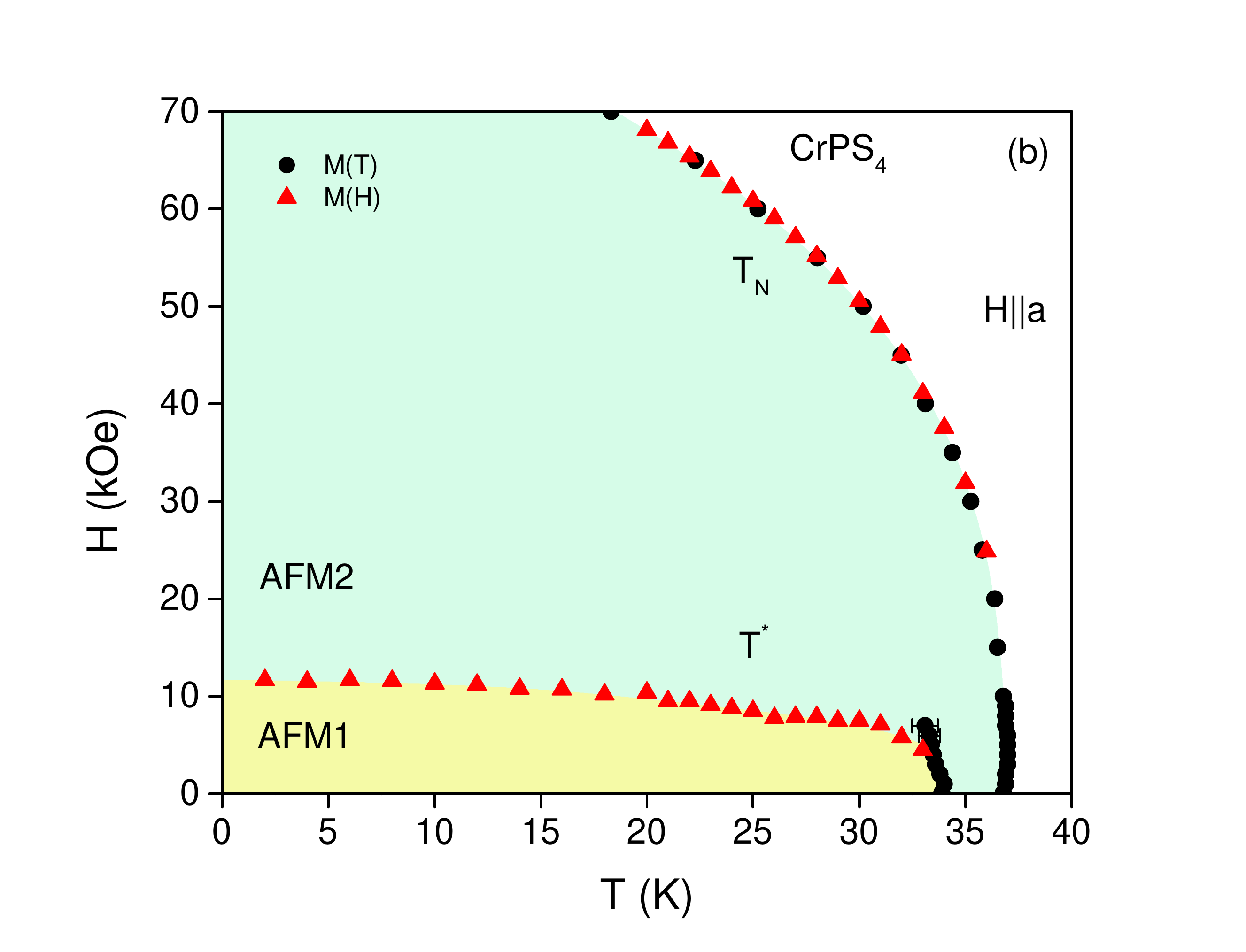}
\includegraphics[angle=0,width=80mm]{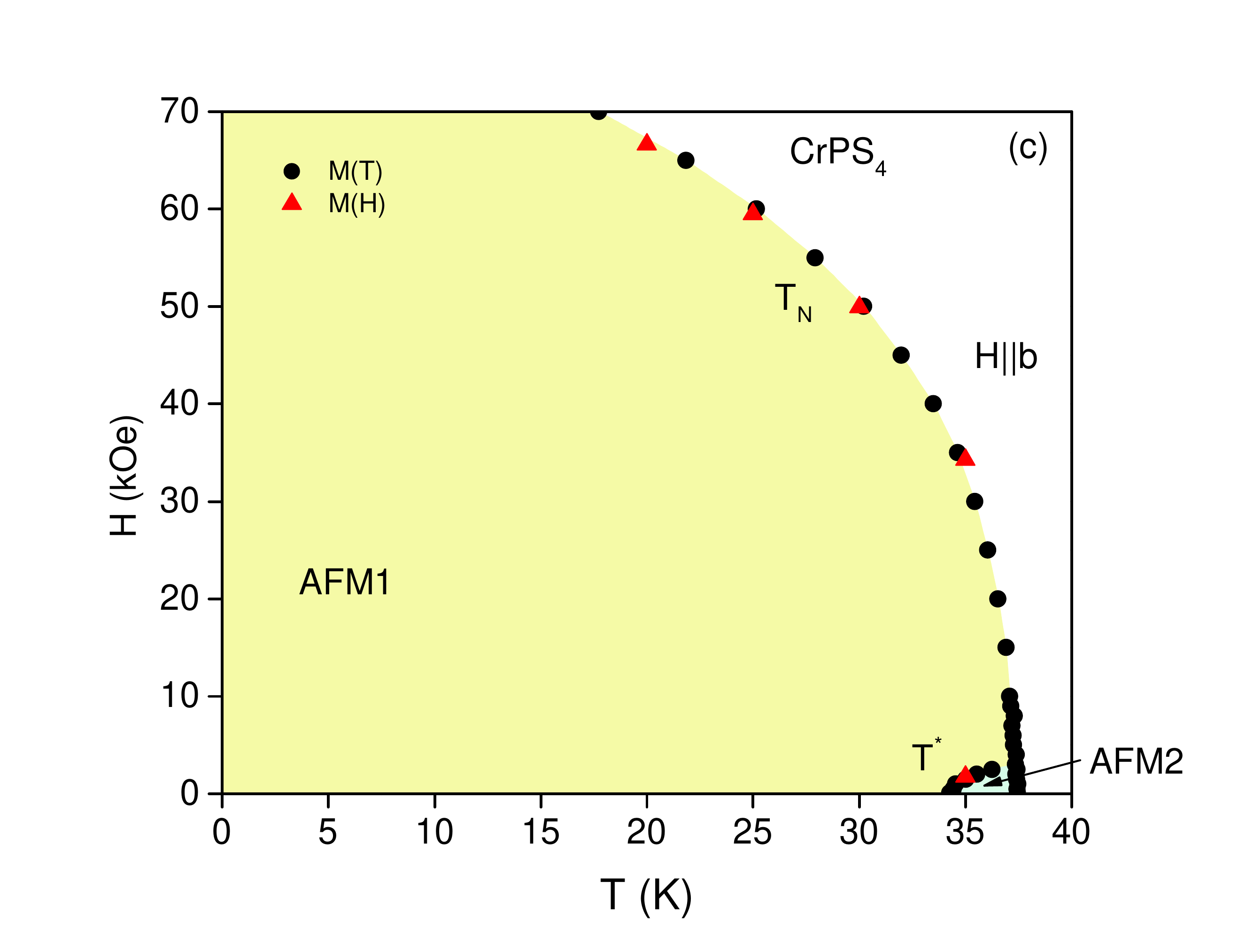}
\end{center}
\caption{(Color online) Anisotropic, ambient pressure $H - T$ phase diagrams constructed based on $M(T)$ and $M(H)$ data. (a) $H \| c$; (b) $H \| a$; (c) $H \| b$. } \label{F6}
\end{figure}

To determine anisotropic $H - T$ phase diagrams, a set of $M(T)$ measurements at constant fields and $M(H)$ measurements at constant temperatures were performed for each magnetic field orientation. A subset of the data together with the criteria used in the analysis of the $M(H)$ data are shown in Figs. \ref{F3}, \ref{F4}, and \ref{F5} [for $M(T)$ data the criteria are the same as above, in Fig. \ref{F2}(b)]. The data for $H \| c$ are consistent with the published results, \cite{lou78a,pei16a,pen20a,din20a} whereas the published data for $H \perp c$  do not specify the in-plane direction and are lacking analysis.\cite{lou78a,pei16a,pen20a} The resulting anisotropic $H - T$ phase diagrams are shown in Fig. \ref{F6}.

For $H \| c$, the $M(T)$ data [Fig. \ref{F3}(a)] show the rapid suppression of the lower transition, $T^*$, with increasing field. The $M(H)$ data, Fig. \ref{F3}(b), show a sharp metamagnetic transition just below 10 kOe at base temperature with its critical field decreasing with increasing temperature. Energetically the $\sim 10$ kOe field needed to stabilize the AFM2 phase is again consistent with the AFM1 and AFM2 states being quite close in energy.

Fig. \ref{F6}(a) shows the position of the phase lines, which are consistent with the literature data. \cite{pei16a,pen20a,din20a} Single crystal neutron scattering studies of CrPS$_4$ in zero field and with magnetic field applied along the $c$ - axis \cite{pen20a,cal20a} identified the AFM1 phase as  the A-type AFM with the magnetic moment along (or slightly canted off) the $c$ axis, whereas in the AFM2 phase the moments are along (or slightly canted off) the $b$-axis. As mentioned above, our low field $M(T)$ data are consistent with the neutron scattering results.

\begin{figure}
\begin{center}
\includegraphics[angle=0,width=80mm]{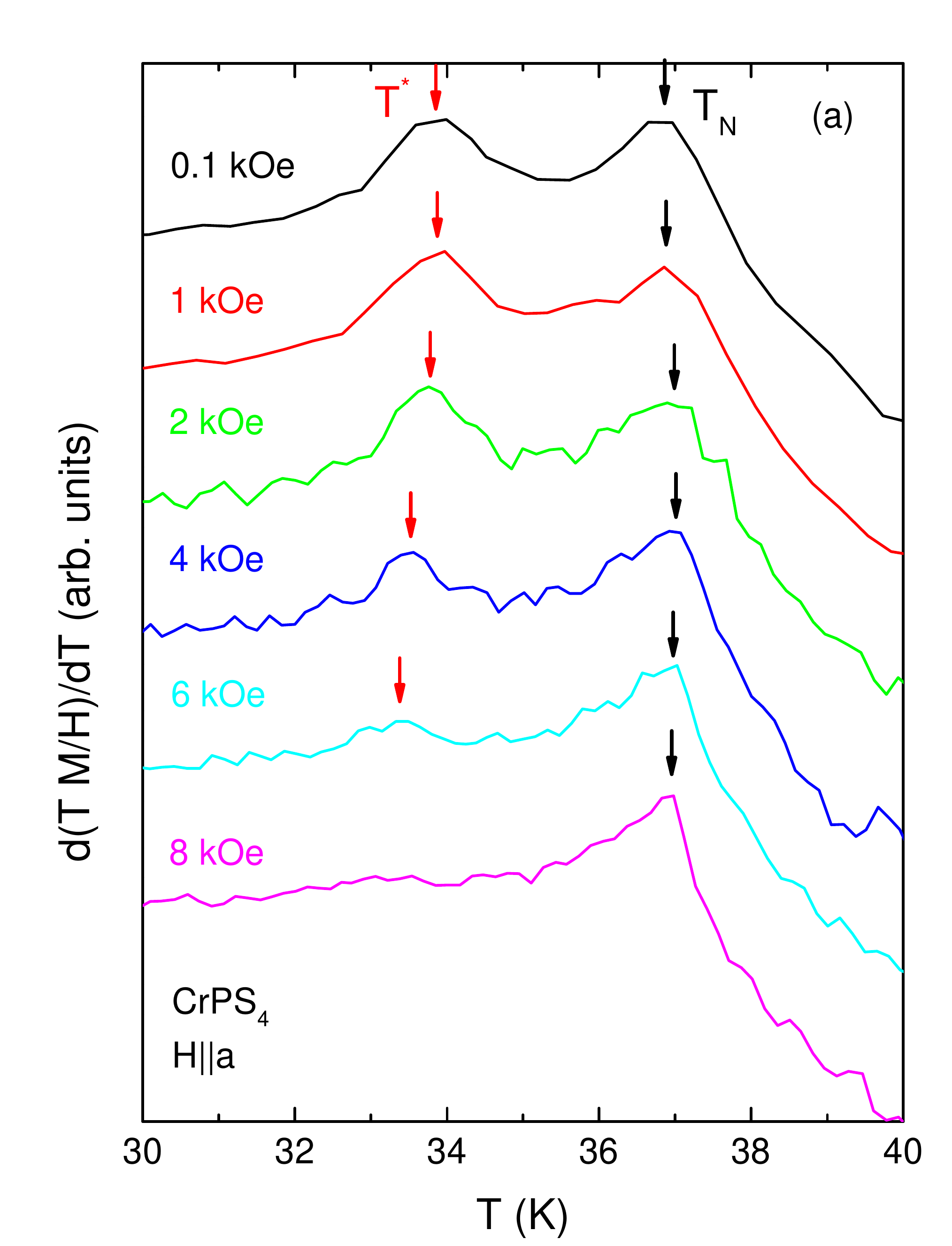}
\includegraphics[angle=0,width=80mm]{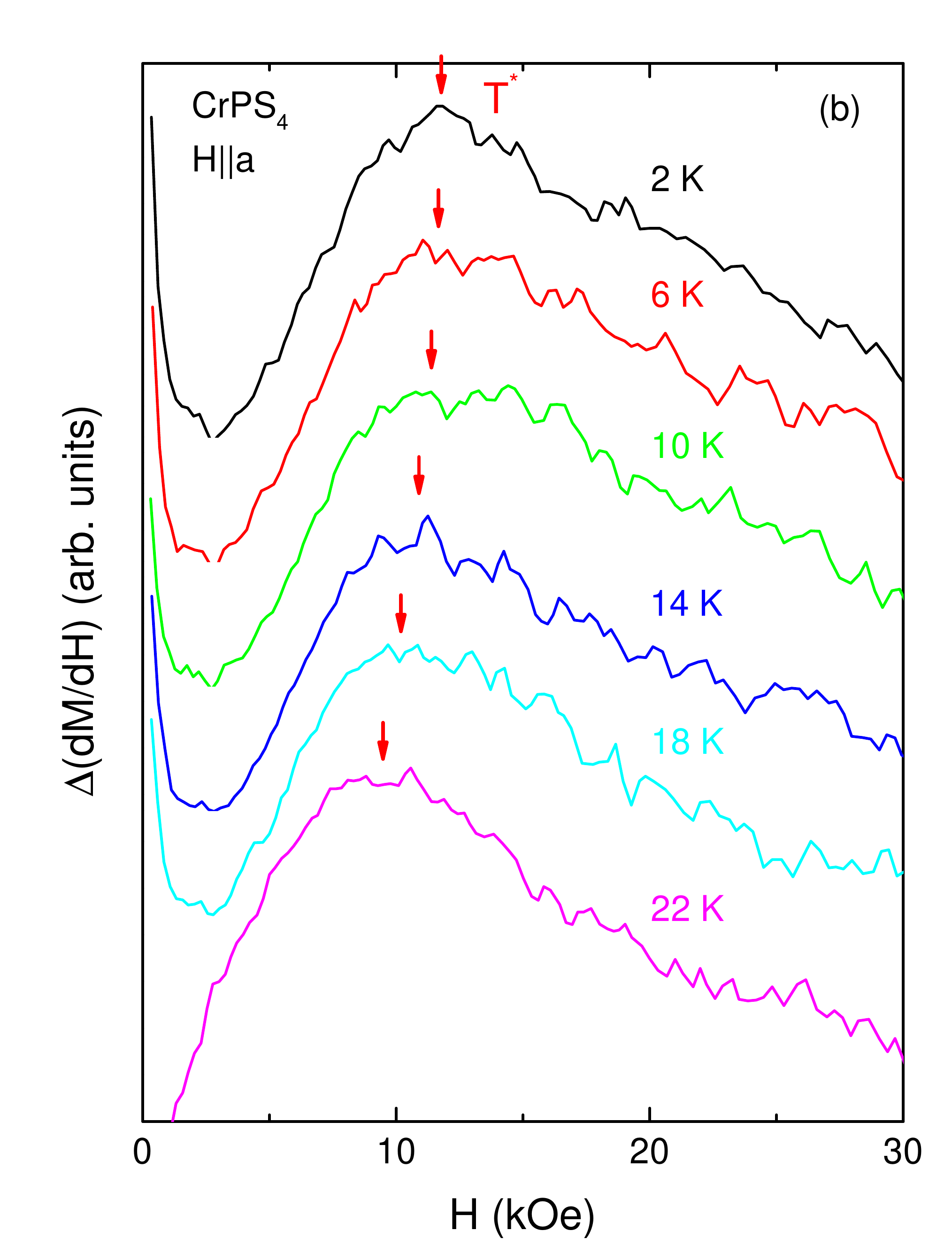}
\end{center}
\caption{(Color online)  (a) Selected examples of the derivatives $d(\chi T)/dT$ vs $T$ of the low field temperature-dependent magnetization data taken for $H \| a$; (b) selected examples of the derivatives $dM/dH$ vs $H$ of the $H \| a$ $M(H)$ data with  linear background subtracted.   Arrows mark the positions of the transitions. In both panels the data are shifted vertically for clarity.} \label{FA1}
\end{figure}

For $H \| a$, the feature in $M(T)$ associated with $T^*$ fades out without significant shift with increasing $H$ and disappears below $\sim 8$~kOe [see Figs. \ref{F4}(a), \ref{FA1}(a)]. $T^*$ is detected in the $dM/dH$ vs $H$ data as a rather subtle feature, a broad maximum [see Figs. \ref{F4}(b), \ref{FA1}(b)]. Its position changes by less than 5 kOe between 2 and 30 K. The $T_N$ feature in both $M(T)$ and $M(H)$ data sets is rather clear. The difference between the signatures of the low temperature $T^*$ lines for $H \| c$ and $H \| a$ [Figs. \ref{F6}(a) and  \ref{F6}(b)] can be understood by the energetics driving the metamagnetic transitions.  For $H \| c$ there is a clear energy difference between the AFM1 and AFM2 states since AFM1 has the moment $\sim$ along the applied field direction and AFM2 has the moments $\sim$ perpendicular to the applied field direction.  For $H \| a$, both the AFM1 and AFM2 states have the moments $\sim$ perpendicular to the applied field direction.  It is possible that the fact that for the AFM1 state the moments are reported to be $\sim 9.5^{\circ}$ degrees off of the $c$-axis (toward the $a$-axis) whereas for the AFM2 state the moments are along the $b$-axis gives rise to a small difference between these states.  From our data it seems likely that the transition from AFM1 to AFM2 for $H \| a$  is more gradual and broad than that for $H \| c$.

For $ H\| b$ [Fig. \ref{F5}(a)], in low fields, the $T^*$ feature rapidly moves towards $T_N$. It is seen in the raw low field $M(H)$ data (Fig. \ref{F5}(b), lower inset) and much more clearly as a sharp low field peak in the $dM/dH$ vs $H$ data at 35 K (Fig. \ref{F5}(b), upper inset) but  is absent in the $M(H)$ data taken at other temperatures. This rapid disappearance of the AFM2 phase can be seen clearly in Fig. \ref{F6}(c).  Again, this behavior can be understood with rather simple energetic arguments.  For applied field along the $b$-axis, the AFM2 state (with moments $\sim$ along the $b$-axis) will become less and less energetically favorable  as field increases, with the AFM1 state with its moments $\sim$ perpendicular to the applied field becoming more favorable.  As a result, the AFM2 state is reduced to a small region or bubble spanning the $T_N$ to $T^*$ temperature range.

The reason for such in-plane anisotropy (the difference between $H \| a$ and $H \| b$) requires further investigation, in particular, whether the structural motif of quasi-1D chains of CrS$_6$ octahedra interconnected along the $a$-axis is of importance. This said, it is curious that in all three phase diagrams, the enveloping $T_N$ phase line between the paramagnetic and magnetically ordered phases is practically the same without notable anisotropy.

\subsection{Effects of pressure on magnetic ordering temperatures}

\begin{figure}
\begin{center}
\includegraphics[angle=0,width=80mm]{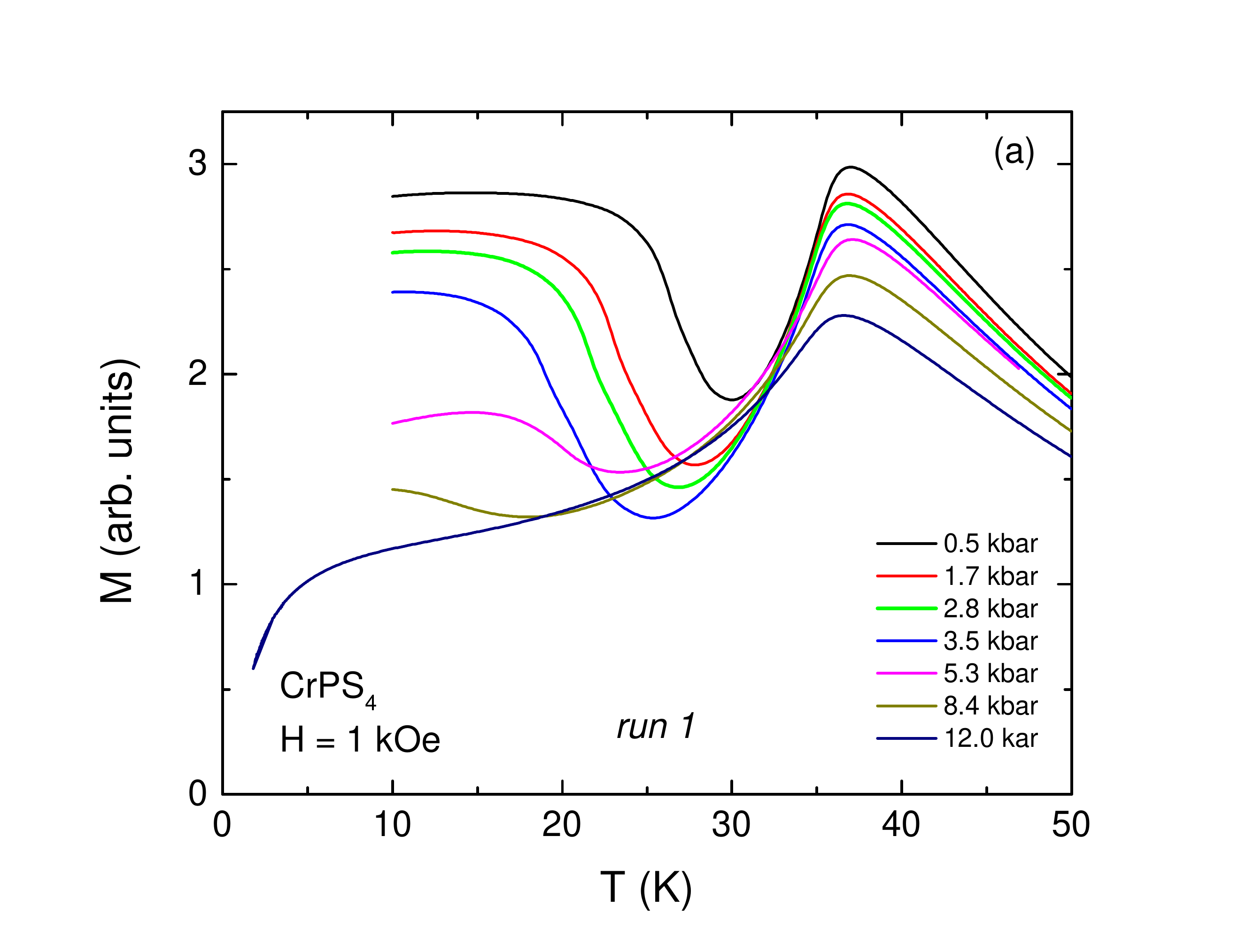}
\includegraphics[angle=0,width=80mm]{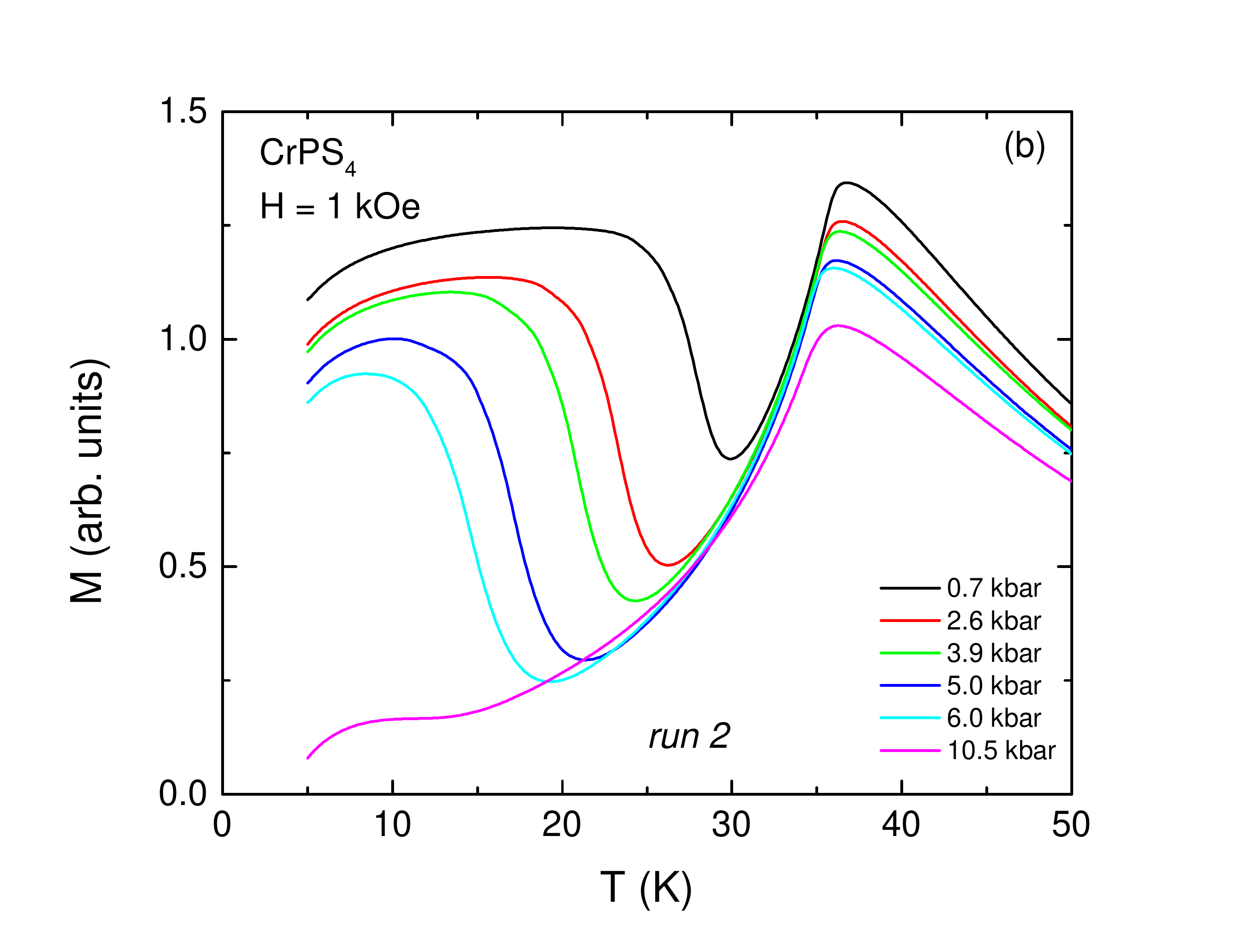}
\includegraphics[angle=0,width=80mm]{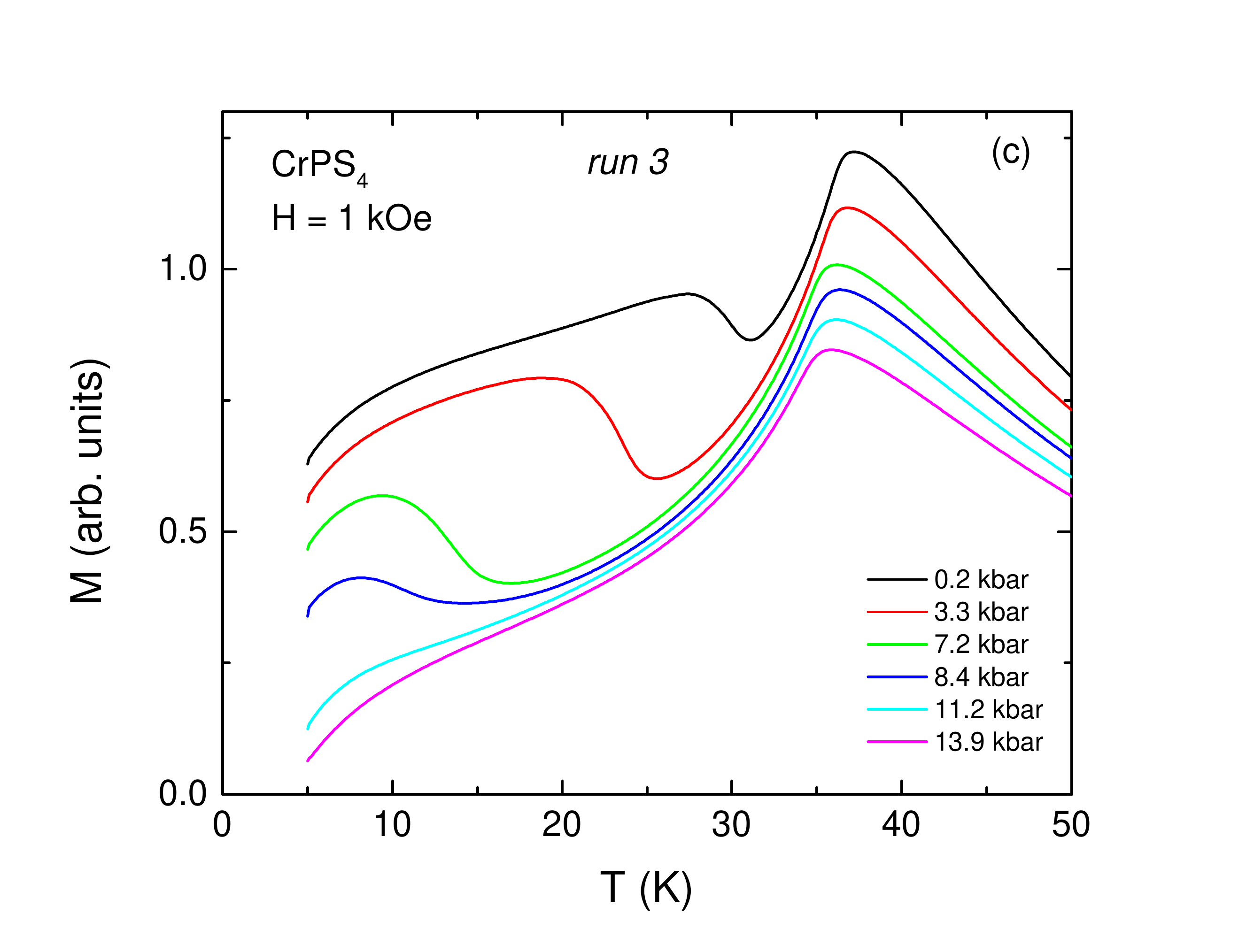}
\end{center}
\caption{(Color online) Temperature dependent magnetization measurements under pressure (runs 1, 2 and 3). The sample orientations in all runs were close to $H \| b$ before application of pressure. } \label{F7}
\end{figure}

Examples of the temperature dependent magnetization measured for the sample orientation close to $H \| b$ at different pressures are shown in Fig. \ref{F7}. Altogether, three magnetization runs under pressure were performed, and although the sample was apparently more misaligned in run 3 (cf. Fig. \ref{F2}), the overall behavior and the measured effect of pressure on the transition temperatures is very similar. Both transition temperatures, determined from $d(\chi T)/dT$, decrease under pressure but with clearly different rates: the lower, $T^*$, is suppressed much faster than the higher, $T_N$. It is noteworthy that in all of these runs (to greater or lesser extents) the size of the low temperature magnetization value in the AFM1 state (i.e. below $T^*$) is decreasing relative to the local maximum at $T_N$,  and indeed the increase-in-magnetization-upon-cooling feature associated with the lower transition  can not longer be resolved for pressures higher than $\sim 10.5$ kbar (see figure \ref{F10} below). This decrease in the lowest temperature magnetization value suggests that the direction of the ordered moment in the AFM1 state (at low temperatures) may well be changing with pressure.

\begin{figure}
\begin{center}
\includegraphics[angle=0,width=120mm]{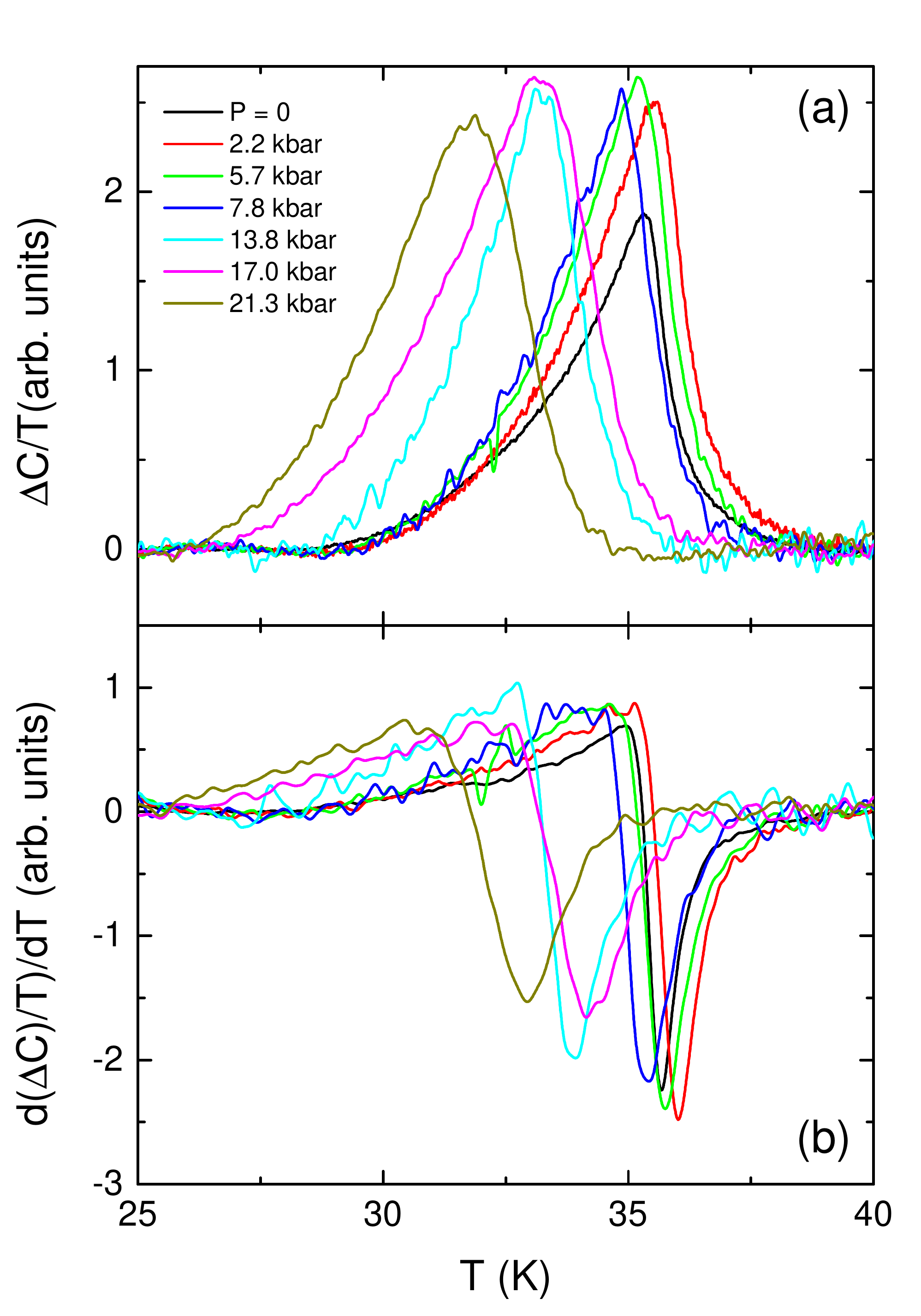}
\end{center}
\caption{(Color online) (a) Enlarged view of the anomalous contribution to the specific heat data, $\Delta C/T$, around $T_N$;  (b) corresponding temperature derivatives, $d(\Delta C)/dT$ .} \label{F8}
\end{figure}

To study the phase diagram of CrPS$_4$ to higher pressures, we conducted specific heat measurements up to 21.3 kbar. In Fig. \ref{F8} (a) we show an enlarged view of the anomalous contribution to the specific heat data, $\Delta C/T$, around $T_N$. $\Delta C/T$ was obtained by subtracting a background contribution from the bare specific heat data. The background contribution was estimated by performing a third-order-polynomial fit  to the $C/T$ data below and above the phase transition at $T_N$, i.e., for $20~\textrm{K} \leq T \leq 25~\textrm{K}$ and $40~\textrm{K} \leq T \leq 45~\textrm{K}$. The so-obtained  $\Delta C/T$ data in Fig. \ref{F8} (a) show clear anomalies that are associated with the high-temperature phase transition at $T_N$. The feature shifts to slightly lower temperatures upon increasing pressure and remains almost unchanged in size except for the $P = 0$ data. This first data point was taken with the lock-nut being hand-tight, which possibly caused inhomogeneous pressure at low temperatures \cite{xia20a} and some thermal decoupling of the sample assembly from the thermal bath (frozen medium). To determine the transition temperature $T_N$, we refer to the positions of the minima in the temperature-derivative of the data, $d( \Delta C/T)/dT$, shown in Fig. \ref{F8} (b). Correspondingly, we infer that $T_N$ is lowered from $\sim 36$ K to $\sim 33$ K by changing pressure from $P = 0$ to 21.3 kbar. As shown in Fig. \ref{F10} (below), the $T_N(P)$ data inferred from magnetization and specific heat measurements agree well with each other.

\begin{figure}
\begin{center}
\includegraphics[angle=0,width=120mm]{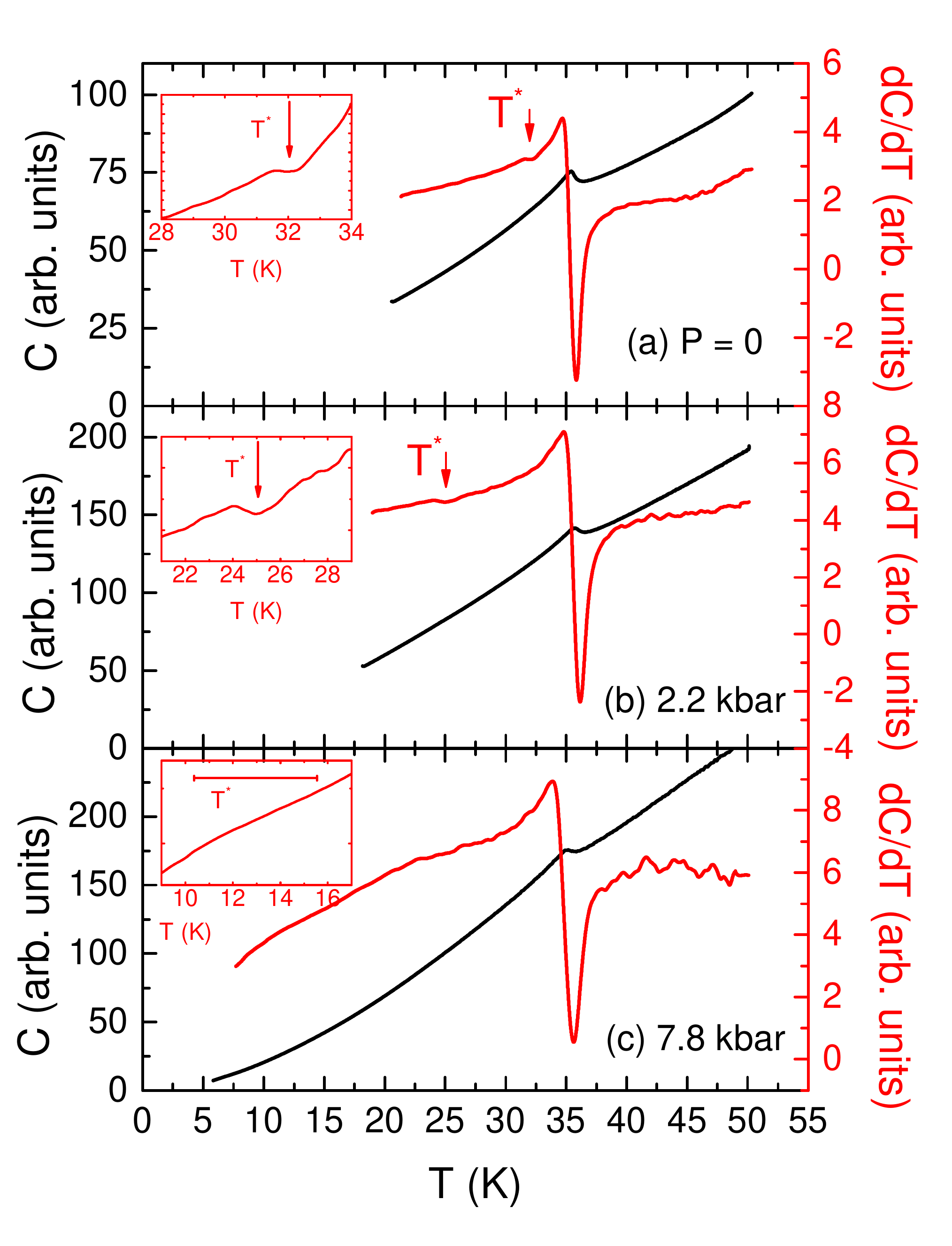}
\end{center}
\caption{(Color online)  $C(T)$ data without background correction (left axis) and the temperature derivatives of the $C(T)$ data (right axis) for (a) $P = 0$, (b) 2.2 kbar,  and (c) 7.8 kbar. Insets in panels (a) and (b) show $dC/dT$ derivatives near the $T^*$ transition on enlarged scales. Inset in panel (c) shows the $dC/dT$ data on an enlarged scale in the temperature range where $T^*$ is expected. The horizontal bar marks $T^*$ values expected for $P = 7.8$~kbar from magnetization data and from extrapolation of two $T^*$ points in heat capacity data.} \label{F9}
\end{figure}

In Fig. \ref{F1} we presented $C_p(T)$ data measured via an adiabatic relaxation technique. The feature associated with $T^*$ is small but resolvable. This said, tracking this feature as a function of pressure is challenging. To discuss the signatures of the phase transition at $T^*$ in the specific heat data collected in the pressure cell, we show in Fig. \ref{F9} data sets of the $C(T)$ data without background correction (left axis) and the temperature derivative of the $C(T)$ data (right axis) for $P = 0$ (a), 2.2 kbar (b) and 7.8 kbar (c). For all pressures, the previously discussed feature at $T_N$ is clearly resolved in $C(T)$ and $dC/dT$. In addition, for $P = 0$ and 2.2 kbar, there is a subtle feature in the specific heat data below $T_N$ which is only barely visible in the raw data. However, this feature gives rise to a small but resolvable minimum in $dC/dT$ (see arrows), which we associate with the transition at $T^*$.  The $T^*$ temperatures determined from $C(T,P)$ for low pressures are consistent with the ones inferred from the $M(T,P)$ data (see figure \ref{F10}). For even higher pressures, e.g., for 7.8 kbar [see Fig. \ref{F9} (c) and inset], we are unable to resolve any feature associated with $T^*$ down to 5 K. We note that for 7.8 kbar, the magnetization data indicates $T^* \approx 15.5$ K. Thus, it is likely that the absence of a resolvable specific heat feature at $T^*$ for higher pressures is related to a decrease of the associated entropy and broadening of the signature of this transition (also observed in magnetization measurements) as the phase transition is suppressed towards lower temperatures.

The data for the pressure dependence of the two transition temperatures, $T_N$ and $T^*$, from magnetization and heat capacity are combined on the $P - T$ phase diagram in Fig. \ref{F10}. The data from different runs and experimental techniques are consistent. Both transition temperatures, $T_N$ and $T^*$,  decrease under pressure in a linear fashion but with very different rates. Whereas the pressure derivatives of $T_N$ are small, $dT_N^M/dP = - 0.07(1)$ K/kbar from magnetization measurements and $dT_N^C/dP = - 0.14(2)$ K/kbar from heat capacity, $T^*$ is suppressed significantly faster: $dT^{*M}/dP = - 1.97(5)$ K/kbar from magnetization measurements, so we extrapolate that  by $\sim 15-16$ kbar the AFM1 phase is completely suppressed.  It has to be noted that the extrapolation of $T^*(P)$ to $P = 0$ results in slightly lower values than those measured by specific heat and magnetic susceptibility at ambient pressure, whereas the $T_N$ data are consistent. A possible explanation is that the first order phase transition at $T^*$ occurs at temperatures when the pressure medium is already frozen. Depending on the details of the lattice parameters change at $T^*$, this might cause small non-hydrostatic pressure component, resulting in a consistent, small $T^*$ shift at each pressure point. This said, we believe that the effect is small and obtained $T^*(P)$ behavior represents the intrinsic properties rather accurately.

\section{Summary}

Specific heat and comprehensive anisotropic magnetization measurements present thermodynamic evidence of two magnetic transitions in CrPS$_4$, at $\sim 34$ K and $\sim 37$ K in zero applied field. Although the $H - T$ phase diagrams for $H \| c$ and $H \| a$ appear to be quite similar,  the AFM1 $\rightarrow$ AFM2 transition on increase of magnetic field is a sharp spin-flop transition in the former case in contrast to being broad and continuous in the latter. The $H - T$ diagram for $H \| b$ appears to be very different: the AFM2  phase that occupies a majority of the phase space for two other orientations is present here only in a small bubble in the high temperature - low field edge of the phase diagram. All of these differences are consistent with the AFM1 phase having the Cr moments aligned within $\sim 10^{\circ}$ of the $c$-axis and the AFM2 phase having the Cr moments aligned along the $b$-axis, as preiously reported.

Hydrostatic pressure up to $\sim 20$ kbar has very small effect on the $T_N$; however, the $T^*$ is suppressed at a high rate, so that by $\sim 15-16$ kbar the AFM1 phase ceases to exist.

Clearly, microscopic measurements on single crystals in an applied field and under pressure are desired to confirm (or refute) the conjectures based on thermodynamic measurements. Additionally, we hope that these results will instigate further theoretical and band-structural studies of magnetic properties of CrPS$_4$ in magnetic field, under pressure, and with stress/strain, in particular, to understand which interactions are responsible for the observed drastic difference in $T_N$ and $T^*$ behavior under pressure. 

Another question that requires further consideration is the extent of similarity between different Cr - based vdW materials. At a first glance, there is a substantial difference between CrPS$_4$, an antiferromagnet with an additional spin orientation transition, and e.g. ferromagnetic Cr$_2$Ge$_2$Te$_6$ and Cr$_2$Si$_2$Te$_6$. \cite{sel20a,cas15a} Still, more, detailed studies are needed to clarify this issue.

\begin{acknowledgments}

Work at the Ames Laboratory was supported by the U.S. Department of Energy, Office of Science, Basic Energy Sciences, Materials Sciences and Engineering Division. The Ames Laboratory is operated for the U.S. Department of Energy by Iowa State University under contract No. DE-AC02-07CH11358. T.J.S. was supported by  Center for Advancement of Topological Semimetals, an Energy Frontier Research Center funded by the U.S. Department of Energy Office of Science, Office of Basic Energy Sciences, through the Ames Laboratory under its Contract No. DE-AC02-07CH11358.  E.G. and T.J.S. were funded, in part, by the Gordon and Betty Moore Foundation’s EPiQS Initiative through Grant No. GBMF4411. 

\end{acknowledgments}

\appendix*

\section{Rietveld refinement of the powder x-ray data}
\label{AA}

\begin{figure}
\begin{center}
\includegraphics[angle=0,width=140mm]{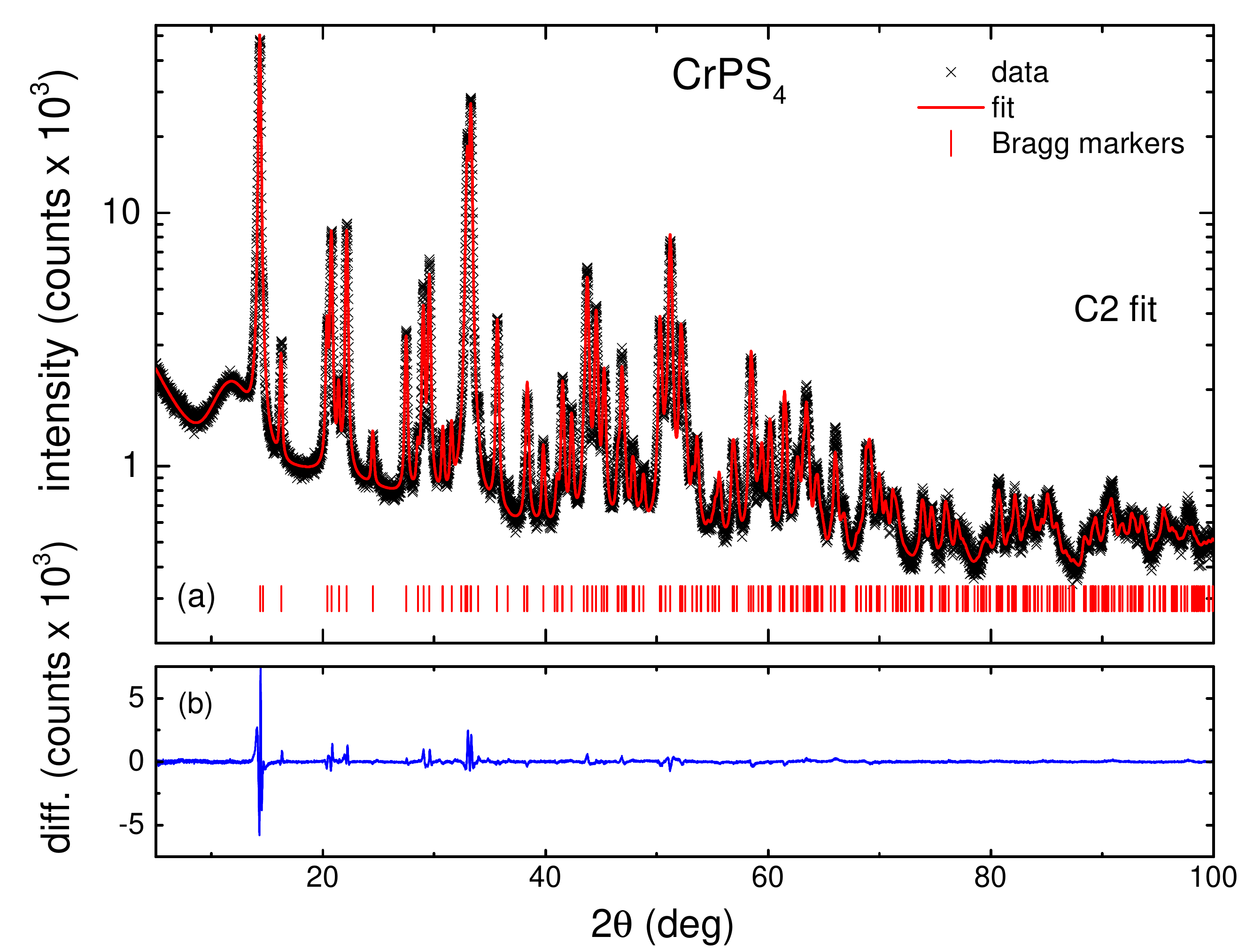}
\end{center}
\caption{(Color online) (a) x-ray diffraction pattern for CrPS$_4$ shown on a semi-log plot. Symbols - data, line - fit to the $C2$ space group, vertical bars - Bragg markers of the peaks; (b) difference between the data and the fit shown on a  linear plot.} \label{FAP1}
\end{figure}

\begin{figure}
\begin{center}
\includegraphics[angle=0,width=140mm]{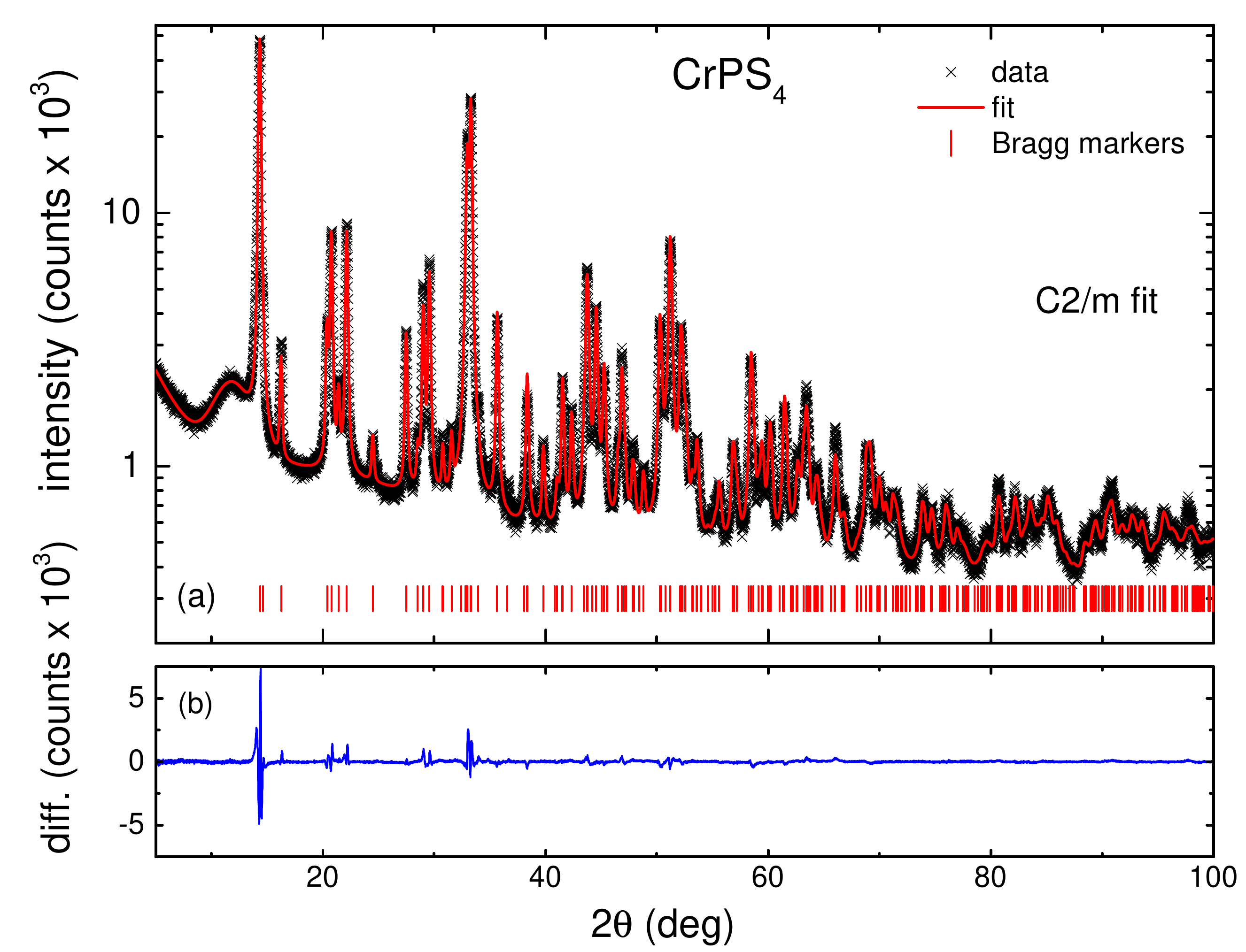}
\end{center}
\caption{(Color online) (a) x-ray diffraction pattern for CrPS$_4$ shown on a semi-log plot. Symbols - data, line - fit to the $C2/m$ space group, vertical bars - Bragg markers of the peaks; (b) difference between the data and the fit shown on a  linear plot.} \label{FAP2}
\end{figure}

Two fits of the powder x-ray diffraction pattern, one with the space group {\it C2}, as reported in Refs. \onlinecite{die77a,lou78a,pen20a,cal20a} and another, with the space group {\it C2/m}, as reported in Refs. \onlinecite{tof77a,pei16a,sus20a} are shown in  figures \ref{FAP1} and \ref{FAP2} respectively. The  lattice parameters and unit cell angles resulting from the fits are listed in Table \ref{T1}.

\begin{table*}[h]

\begin{ruledtabular}

\caption{Lattice parameters and unit cell angles obtained from the Rietveld refinements of the CrPS$_4$  powder x-ray diffraction data using two different space groups. Note: listed error are taken from the software output. \\}{\label{T1}} 
\begin{tabular}{cccccc}
Space group&$a$ (\AA)&$b$ (\AA)&$c$ (\AA)&$\alpha = \gamma$ (deg)&$\beta$ (deg)\\ \hline
{\it C2}&10.8597(6)&7.2500(4)&6.1427(3)&90&91.893(2)\\
{\it C2/m}&10.8575(7)&7.2485(4)&6.1418(3)&90&91.895(2)\\

\end{tabular}
\end{ruledtabular}
\end{table*}

The values of the lattice parameters and the angle $\beta$ are consistent with the published values. \cite{tof77a,die77a,lou78a,pei16a,pen20a,cal20a,sus20a} From the figures and from very close goodness of fit (3.36 for {\it C2} fit vs 3.44 for {\it C2/m} fit) and weighted profile residual, $R_{wp}$,  (8.956 for {\it C2} fit vs 9.181 for {\it C2/m} fit) values  it is clear that the powder xrd data from the laboratory instrument do not allow to distinguish between these two possibilities. 

This existing ambiguity in reported crystal structure for CrPS$_4$ gives rise to the possibility that the studied samples of CrPS$_4$ are  actually a mixture / intergrowth of the {\it C2} and {\it C2/m} crystallographic phases.  If both the {\it C2} and {\it C2/m} regions were above some minimum size, then there could, hypothetically be two magnetic transitions, $T_N$  and $T^*$, each associated with their own distinct phase. Whereas this is a possibility, in our opinion, neither results of this work, nor the existing publications on the physical properties of CrPS$_4$, contain any experimental evidence supporting such hypothesis, making it highly unlikely.

\clearpage

\clearpage

\clearpage

\clearpage

\clearpage

\clearpage

\clearpage

\clearpage

\clearpage

\clearpage

\clearpage

\clearpage

\clearpage

\begin{figure}
\begin{center}
\includegraphics[angle=0,width=120mm]{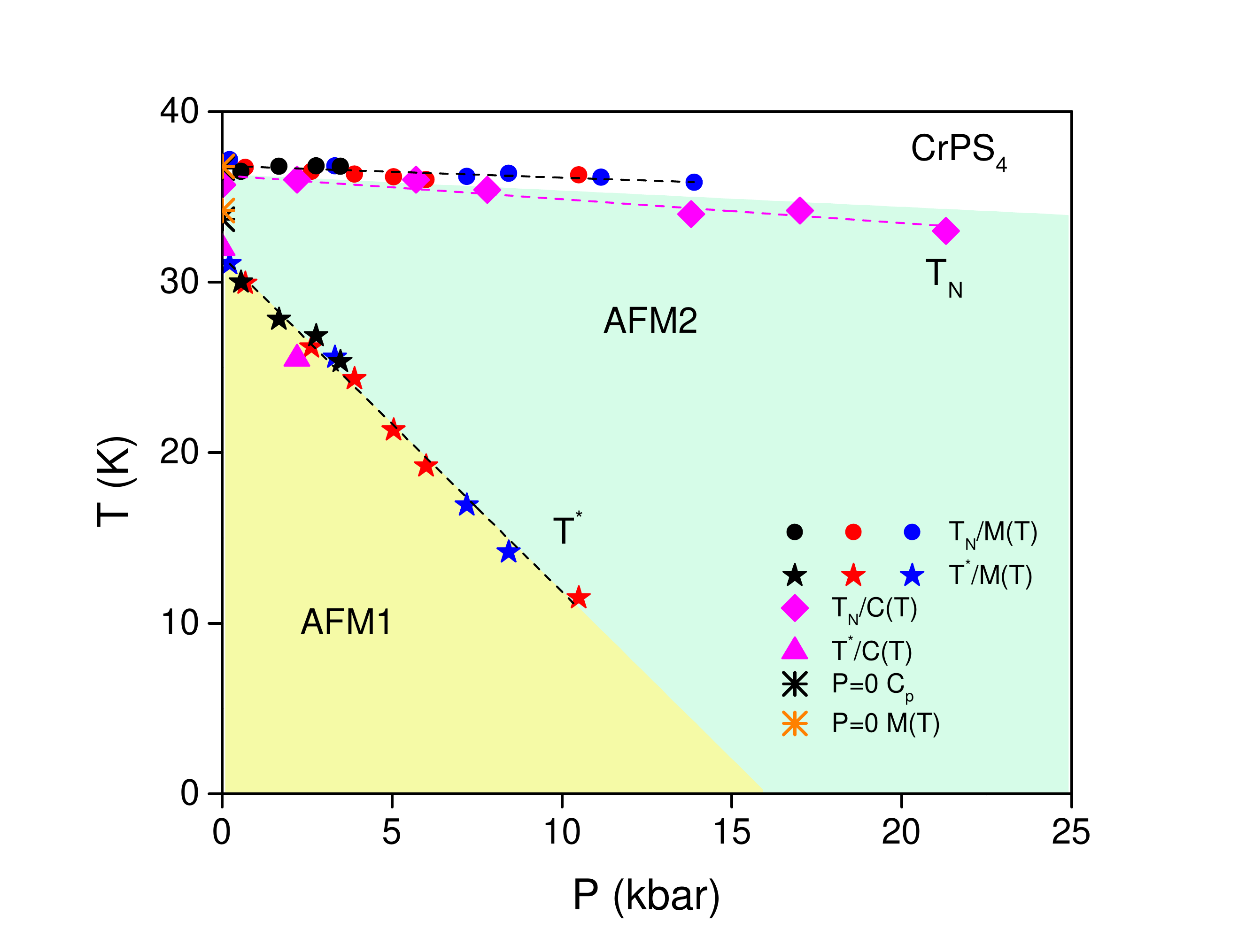}
\end{center}
\caption{(Color online)  $P - T$ phase diagram of CrPS$_4$ based on magnetization (different colors correspond to different runs) and heat capacity measurements under pressure. Black and orange asterisks are the $P = 0$ transition temperatures from specific heat and $H \| b$, 1 kOe magnetic susceptibility measurements respectively.  Dashed lines are linear fits of $T_N(P)$ and  $T^*(P)$ data from magnetization runs, as well as  $T_N(P)$  data from heat capacity measurements. Color backgrounds highlighting different phases are guide to the eye and are  based on extrapolation of the experimental phase lines.} \label{F10}
\end{figure}

\clearpage

\clearpage

\end{document}